\documentclass[preprint, 12pt,a4paper,onecolumn,groupedaddress,preprintnumbers,nofootinbib,floatfix]{revtex4-1}
\usepackage[top=3cm, bottom=2.1cm, left=2cm, right=2cm]{geometry}
\pdfoutput=1

\usepackage{comment}

\usepackage{graphicx}
\usepackage{url}
\usepackage[bookmarks, pagebackref=false]{hyperref}
\usepackage[usenames,dvipsnames]{xcolor}
\usepackage{dcolumn}
\usepackage{bm}
\usepackage{bbm}
\usepackage{amsmath,amssymb,amsfonts}
\usepackage{color}
\usepackage{hyperref}
\usepackage[Symbolsmallscale]{upgreek}
\usepackage{dsfont}
\usepackage[all]{xy}
\usepackage{pstricks}
\usepackage{dsfont}%
\usepackage{mathtools}
\usepackage{placeins}
\usepackage{enumerate}
\usepackage{cases}
\usepackage{placeins}
\usepackage{xspace}
\usepackage{cancel} 
\usepackage{slashed}
\usepackage{subfigure}
\usepackage{natbib}
\hypersetup{
    bookmarks=true,         
    unicode=false,          
    pdftoolbar=true,        
    pdfmenubar=true,        
    pdffitwindow=false,     
    pdfstartview={FitH},    
    pdftitle={Directional Detection},    
    pdfauthor={Author},     
    pdfsubject={Subject},   
    pdfcreator={Creator},   
    pdfproducer={Producer}, 
    pdfkeywords={keyword1} {key2} {key3}, 
    pdfnewwindow=true,      
    colorlinks=true,       
    linkcolor=magenta,          
    citecolor=magenta,        
    filecolor=magenta,      
    urlcolor=cyan           
}
\usepackage{ulem}

\newcommand{\be}{\begin{equation}}
\newcommand{\ee}{\end{equation}}
\newcommand{\beq}{\begin{eqnarray}}
\newcommand{\eeq}{\end{eqnarray}}

\newcommand{\bmp}{\noindent\begin{minipage}{16cm}}
\newcommand{\emp}{\end{minipage}\vskip 7mm} 

\newcommand{\bvec}[1]{{\bf{#1}}}

\def\lsim{\mathrel{\rlap{\lower4pt\hbox{\hskip1pt$\sim$}}
    \raise1pt\hbox{$<$}}}                
\def\gsim{\mathrel{\rlap{\lower4pt\hbox{\hskip1pt$\sim$}}
    \raise1pt\hbox{$>$}}}                

\begin{document}
\title{Velocity Dependent Dark Matter Interactions in Single-Electron Resolution Semiconductor Detectors with Directional Sensitivity}
\author{Matti Heikinheimo}
\email{matti.heikinheimo@helsinki.fi}
\affiliation{Helsinki Institute of Physics and Department of Physics, University of Helsinki 
}

\author{Kai Nordlund}
\email{kai.nordlund@helsinki.fi}
\affiliation{Helsinki Institute of Physics and Department of Physics, University of Helsinki 
}

\author{Kimmo Tuominen}
\email{kimmo.i.tuominen@helsinki.fi}
\affiliation{Helsinki Institute of Physics and Department of Physics, University of Helsinki 
}

\author{Nader Mirabolfathi}
\email{mirabolfathi@physics.tamu.edu}
\affiliation{Department of Physics and Astronomy,
Texas A\& M University}

\begin{abstract}
We investigate the velocity and recoil momentum dependence of dark matter interactions with ordinary matter. In particular we focus on the single-electron resolution
semiconductor detectors, which allow experimental assessment of sub-GeV
dark matter masses. We find that, within a specific mass range depending on
the detector material, the dark matter interactions result in a signal characterized by daily modulation. Furthermore, we find that the detailed structure of this modulation is sensitive to the velocity and momentum dependence of
dark matter interactions. 
We identify the optimal mass range for the prevalence of these effects. 
\end{abstract}


\preprint{HIP-2019-8/TH}

%
\maketitle

\section{Introduction}
Cosmological and astrophysical observations provide overwhelming evidence for the existence of dark matter (DM) consisted of particles beyond the Standard Model (SM). 
The well established paradigm is a Weakly Interacting Massive Particle (WIMP), with an electroweak scale mass $\mathcal{O}(100)$ GeV. Direct detection of the DM particles has been the target of numerous experimental programs~\cite{Akerib:2017kat, Cui:2017nnn, Aprile:2018dbl}. So far these experiments have not led to a consistent discovery, but have provided solid constraints limiting the strength of the interactions between dark and ordinary matter. The current direct detection experiments are most effective around the typical WIMP mass range of $\mathcal{O}(10)-\mathcal{O}(100)$ GeV, while at smaller masses the existing
experimental constraints~\cite{Agnese:2017njq} are less severe. This has motivated an increasing interest to the low mass region $m_{\rm DM}\lesssim 1$ GeV, see e.g. \cite{Battaglieri:2017aum} and the references therein. Since the coherent elastic scattering of solar neutrinos produce signals that mimic those expected from low mass DM interactions, these experiment will eventually become background limited and need to develop methods to mitigate this irreducible background.

A promising method for direct detection of low mass DM particles was presented in~\cite{Kadribasic:2017obi}, based on single-electron resolution semiconductor detectors. Assuming a direct correlation between ionization and defect creation thresholds in semiconductors, it was noted that due to the anisotropic structure of the semiconductor crystals, the ionization threshold can be sensitive to the recoil direction. 
Hence this technique allows for a detection of a daily modulation signal, due to the rotation of the earth with respect to the direction of the DM wind.

The purpose of this paper is to extend the analysis presented in~\cite{Kadribasic:2017obi}, to cover the variety of non-relativistic operators describing the DM-SM scattering. The directional detection in the context of non-relativistic effective theory of DM-SM scattering has been discussed in~\cite{Catena:2015vpa,Kavanagh:2015jma}, where the angular recoil distributions expected for the various effective operators have been described. Recording the full angular recoil spectrum could thus be used to identify the operator characterizing the DM-SM scattering, and would reveal valuable information about the underlying theory of DM. See \cite{Mayet:2016zxu} for an overview of the prospects in directional DM detection.

Our method, however, does not rely on recording the angular differential recoil distribution, but rather in observing the daily modulation in the total integrated event rate, the origin of which is in the directional sensitivity of the threshold energy. Therefore, the question becomes, to which extent is the modulation signal sensitive to the type of the effective DM-SM scattering operator? Qualitatively, the following behavior is to be expected: The amplitude of the daily modulation signal grows towards lighter DM mass, as the threshold energy for creating the electron-hole pair becomes more significant in comparison to the average kinetic energy available for the recoil. On the other hand, as the DM mass is decreased, the angular recoil spectrum becomes peaked at forward direction for all effective operators. This effect is due to the recoil kinematics {\textit i.e.} if the recoil angle is large, only a small fraction of the kinetic energy of the incoming DM particle is available for the recoil energy of the nucleus.

In the following we will quantify the above assertions, and provide analytic
formulas for the relevant event rates. Based on our results, we establish  the
following general picture {\textit i.e.} in the small mass region where the daily
modulation amplitude is readily observable, the recoil dynamics are largely
insensitive to the effective DM-SM scattering operator. As the DM mass increases,
the different recoil dynamics cause an increasing difference in the
angular distribution. However, this effect becomes masked by the fact that
the amplitude of the daily modulation signal
quickly decreases as the DM mass increases, thus the effect becomes less observable.
We will thus arrive to the conclusion, that
depending on the detector material, there is a range of DM masses
below $\mathcal{O}(1)$~GeV, wherein the direct detection technique described in~\cite{Kadribasic:2017obi} is valid.
At the lower edge of this mass range, the shape of the modulation signal is practically blind to the details of the underlying theory of DM-SM scattering,
but for increasing masses the shape of the modulation signal strongly depends on the velocity dependence of the underlying DM-SM interactions.
The most important parameter for the efficiency of this method is the DM mass.

The paper is organized as follows: in section \ref{sec:basics} we introduce the kinematics of the scattering event and the non-relativistic operators. In section \ref{sec:directional} we describe the directional sensitivity of our detector concept and the resulting daily modulation signal, and we conclude in section \ref{sec:conclusions}. The analytic formulas for the angular event rates are given in appendix \ref{app:radontrans}.

\section{Basic formulas}
\label{sec:basics}
Consider scattering of a dark matter particle with a nucleus. Denote the DM
velocity by $\bvec{v}$\footnote{Throughout this paper we will denote three-vectors, such as $\bvec{v}$, with a boldface font and their amplitudes with italic, as $v$.}. The double-differential recoil rate per unit detector mass is~\cite{Gondolo:2002np}
\be
\frac{d^2R}{dE d\Omega_q}=\frac{\rho_0}{2\pi m_{\rm{DM}}}
\frac{|{\cal{M}}|^2}{32\pi m_{\rm N}^2 m_{\rm{DM}}^2}\delta(\bvec{v}\cdot\hat{\bvec{q}}-v_{\rm{min}}),
\label{eq:diffrate}
\ee
where $m_{\rm{DM}}$ and $m_{\rm N}$ are the masses of the DM particle and nucleus,
respectively. The local DM density is denoted by $\rho_0=0.3\ {\rm GeV}/{\rm cm}^3$ and the direction
of the recoiling nucleus by the unit vector $\hat{\bvec{q}}$. The squared
scattering matrix element
$|{\cal{M}}|^2$ is summed and averaged over the inital and final spins.
The $\delta$-function imposes the kinematic of the elastic scattering, and
the minimum WIMP speed required to excite a nuclear recoil of energy $E=q^2/2m_{\rm N}$ is
\be
v_{\rm{min}}=\sqrt{\frac{m_{\rm N}E}{2\mu^2_{{\rm{DM,N}}}}},
\ee
where $\mu_{{\rm{DM,N}}}$ is the reduced mass of the DM-nucleus system. The angular differential event rate is obtained from (\ref{eq:diffrate}) by integrating over energy:

\be
\frac{dR}{d\Omega_q}=\int_{E_{\rm{min}}}^{E_{\rm{max}}}\frac{d^2R}{dEd\Omega_q}dE,
\label{eq:directional rate}
\ee
where $E_{\rm min}$ is the threshold energy for creating a detectable recoil event, and $E_{ \rm max}$ is the maximum energy allowed by the event selection of the experiment. If no upper bound is imposed by the detection technique, $E_{\rm max}$ can be taken to infinity, as the convergence of the integral is ensured by the integrability of the DM velocity distribution, to be discussed below. In this paper we will take $E_{\rm max}=\infty$ unless otherwise noted.

To calculate the observable directional event rate in a detector on earth, the distribution of DM velocities
in the galactic halo must be taken into account: the rate in Eq.\eqref{eq:diffrate}
must be integrated over all DM velocities weighted by the distribution $f(\bvec{v})$.
In this work we will use the Standard Halo Model, defined as a truncated Maxwellian distribution
\be
f_{\rm{SHM}}(v)=N_{\rm{e}}^{-1}f_M(v)\Theta(v_e-v),
\label{eq:fSHM}
\ee
where \mbox{$v_{\rm e}=537\ {\rm km}/{\rm s}$} is the escape velocity,
\mbox{$f_M(v)=(2\pi\sigma_v^2)^{-3/2}\exp(-v^2/2\sigma_v^2)$} is the Maxwellian distribution with a standard deviation \mbox{$\sigma_v = v_0/\sqrt{2}$}, the circular speed \mbox{$v_0 = 220\ {\rm km}/{\rm s}$} and the normalization constant is given by
\be
N_{\rm e} = {\rm erf}\left(\frac{v_{\rm e}}{\sqrt{2\sigma_v^2}}\right) -\sqrt{\frac{2}{\pi\sigma_v^2}}v_{\rm e} e^{-\frac{v_{\rm e}^2}{2\sigma_v^2}}.
\ee
Taking all the above together, the angular differential rate becomes
\be
\frac{dR}{d\Omega_q}=\frac{\rho_0}{2\pi m_{\rm{DM}}}
\frac{1}{32\pi m_{\rm N}^2 m_{\rm{DM}}^2} \int\limits_{E_{\rm{min}}}^{E_{\rm{max}}} dE \int d^3v\, |{\cal{M}}|^2f_{\rm{SHM}}(v) \delta(\bvec{v}\cdot\hat{\bvec{q}}-v_{\rm{min}}).
\label{eq:diffrate2}
\ee

The integration over the recoil energy $E$ and the DM velocity $\bvec{v}$ is affected by the fact that the squared matrix
element can in principle depend both on $q$ and $\bvec{v}$. For a systematical analysis,
we consider the non-relativistic effective field theory constructed in~\cite{Fitzpatrick:2012ix}.
The effective field theory operator basis is constructed by imposing the requirement
of Hermiticity together with invariance under Galilean transformations and time reversal. In particular, because of
Hermiticity, the velocity dependence of these operators is only through the
combination
\be
v_\perp^2=v^2-\frac{q^2}{4\mu^2_{{\rm{DM,N}}}},
\ee
which by construction satisfies $\bvec{v}_{\perp}\cdot\bvec{q}=0$.

For our purposes it is sufficient to categorize different interactions in terms
of the velocity and energy dependence they imply for the square of the
averaged matrix element appearing in Eq.~\eqref{eq:diffrate2}. The possible dependences
are~\cite{Kavanagh:2015jma,Fitzpatrick:2012ix,DelNobile:2018dfg}
\be
|{\cal{M}}|^2= a_1 1+a_2q^2+a_3 q^4 +b_1v_\perp^2+b_2 q^2 v_\perp^2+  b_3q^4v_\perp^2+\cdots
\label{eq:effinteractions}
\ee
where the ellipsis stands for operators of higher order in $q^2$ and $a_i, b_i$ are coefficients with mass dimension $-2(i-1)$. In addition to these, it is interesting to consider effects from long rage interactions
mediated by some light field. These will lead to behavior $\sim q^{-4}$.

Hence, in order to probe the full range of different
behaviors due to different interactions, we only need to compute two different integrals
over the velocity distribution. The Radon transform, defined as
\be
\hat{f}(v_{\rm{min}},\hat{q})=\int d^3v\, f(v)
\delta(\bvec{v}\cdot\hat{\bvec{q}}-v_{\rm{min}}),
\label{eq:radontrafo}
\ee
corresponds to the velocity dependence ${\cal{O}}(v^0)$ of the matrix element.
The only other possibility, then, is that the squared matrix element is proportional to the square of the perpendicular velocity $v_\perp^2$, and leads to the transverse Radon transform
\be
\hat{f}^{\rm{T}}(v_{\rm{min}},\hat{q})=\int d^3v\, f(v) v_\perp^2
\delta(\bvec{v}\cdot\hat{\bvec{q}}-v_{\rm{min}}).
\label{eq:tradontrafo}
\ee

The angular differential event rate (\ref{eq:diffrate2}) can then be expanded as
\be
\frac{dR}{d\Omega_q} =\frac{\rho_0}{4\pi m_{\rm{DM}}}\frac{\sigma_0 A^2}{\mu^2_{\rm DM,N}} \int\limits_{E_{\rm{min}}}^{E_{\rm{max}}} dE\, \left( (a_1 + a_2 q^2+\ldots)\hat{f}_{\rm SHM}(v_{\rm{min}},\hat{q})
+(b_1 + b_2 q^2 +\ldots) \hat{f}^{\rm{T}}_{\rm SHM}(v_{\rm{min}},\hat{q})\right),
\label{eq:diffrate_expansion}
\ee
where $A$ is the mass number of the nucleus and $\sigma_0 = 1/(16\pi A^2(m_{\rm DM}+m_{\rm N})^2)$ is a reference DM-nucleon cross section. For the SHM the integral over energy can be performed analytically, and the necessary explicit formulas are provided in the appendix \ref{app:radontrans}.

Notice that the overall normalization of the terms in the expansion (\ref{eq:diffrate_expansion}) must include the corresponding nuclear matrix elements~\cite{Kavanagh:2015jma}. In this work our goal is not to determine these absolute normalizations, but to determine the shape of the resulting observable signal which our detector concept would measure given enough exposure, and whether the shape of the signal, as a function of time, is sensitive to the structure of the underlying operators. Therefore we absorb the normalization of the operators in the coefficients $a_i,b_i$. We also neglect the nuclear form-factors which suppress high energy recoils. In the low-mass region we are considering, the recoil energies are small and the form factors are very close to one. We provide a compilation of necessary general formulas for the event rates, which complement existing literature and are expected to be useful for similar studies within different detector concepts currently under active investigation~\cite{Battaglieri:2017aum}.

\section{Directional Energy Threshold}
\label{sec:directional}

\begin{figure}
\begin{center}
\includegraphics[width=0.49\textwidth]{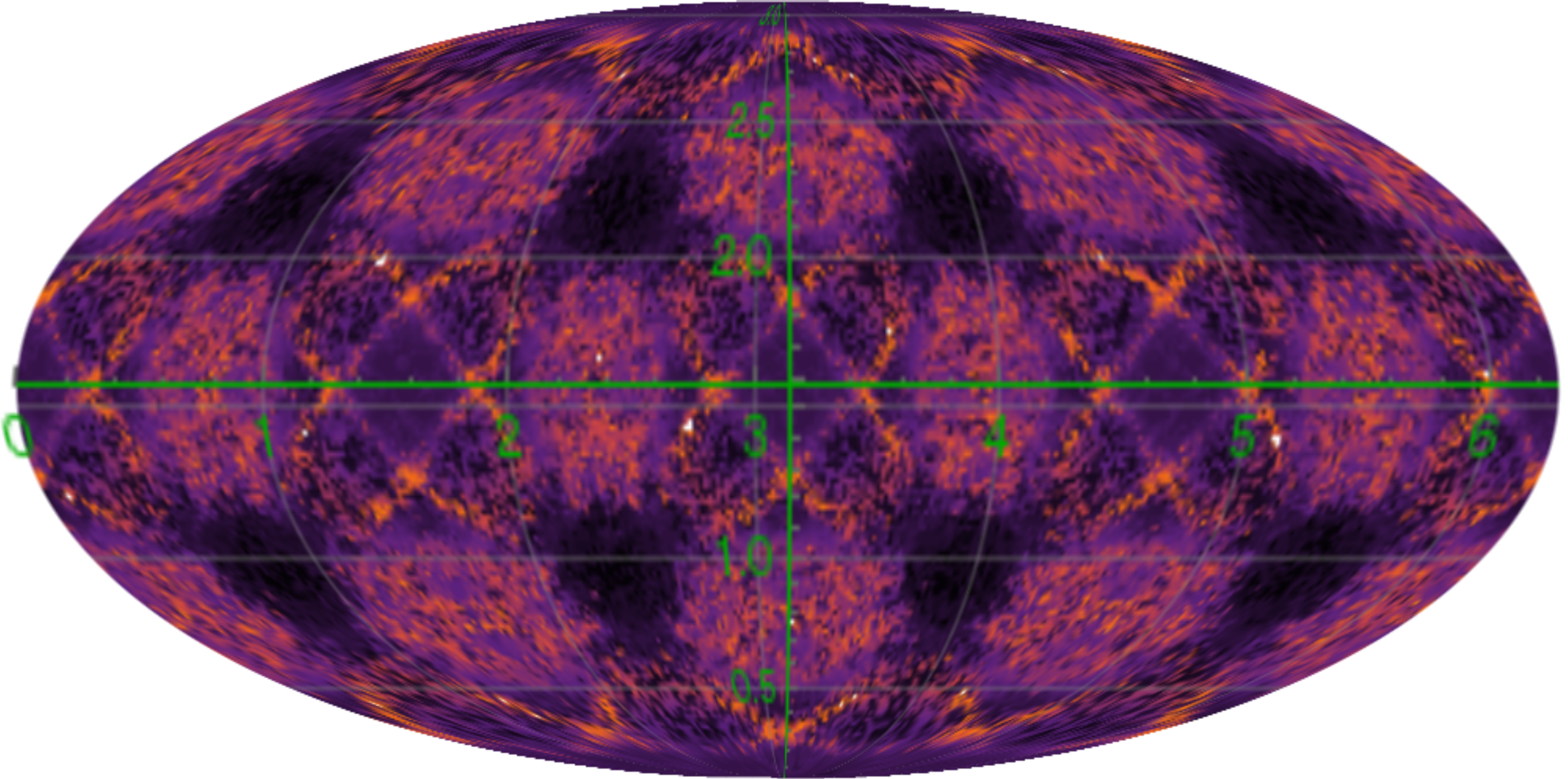}
\includegraphics[width=0.49\textwidth]{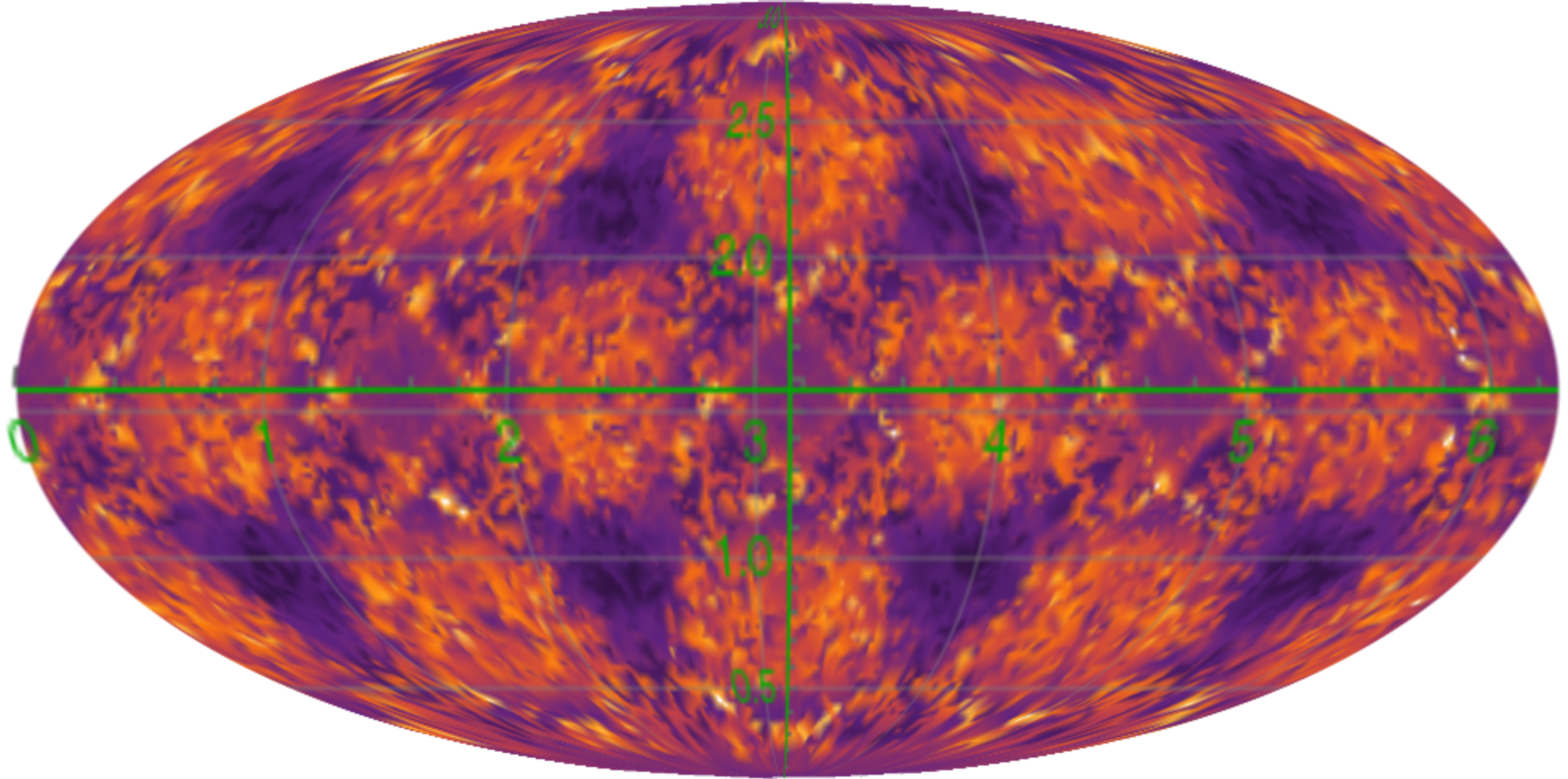}\\
\vspace{0.2cm}
\includegraphics[width=0.4\textwidth]{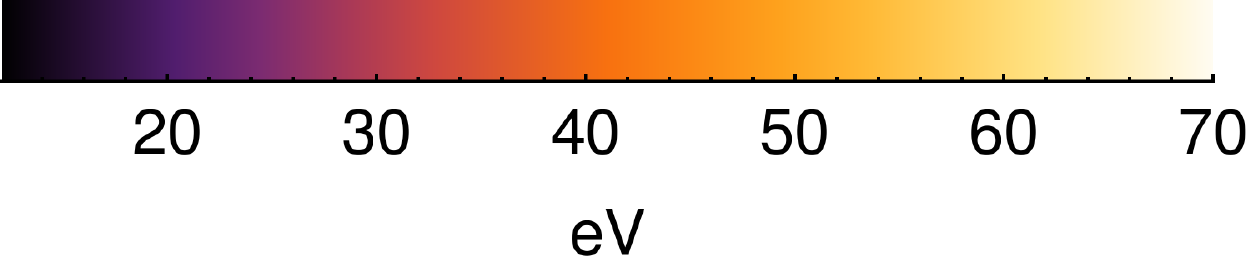}
\caption{The simulated defect creation energy threshold of Germanium (left) ans Silicon (right) as a function of the recoil direction. The recoil energies are given in units of eV.}
\label{fig:GE threshold}
\end{center}
\end{figure}

In semiconductor materials the threshold energy for defect creation is a function of the recoil direction.
The representation of this effect for Germanium and Silicon is shown in figure \ref{fig:GE threshold}. To obtain this figure we have generated a sample of 84936 randomly sampled directions in Germanium and 24155 directions in Silicon, with the corresponding energy thresholds,
utilizing the data from the molecular dynamics simulations \cite{Holmstrom:2008,Nordlund:2006}
carried out in Ref. \cite{Kadribasic:2017obi}. Briefly, Ge and Si atom recoils were
simulated in randomly generated directions in three dimensions. The
directions were selected to give a uniform distribution over solid
angle, i.e. the $\theta$ angle was selected as $\cos^{-1} (1-2u)$
where $u$ is a uniformly distributed random number between 0 and 1
\cite{Press:1992zz}. For each direction, the recoil energy was
increased from 4 eV in 1 eV increments until a stable defect was
produced. Time-dependent density functional theory calculations \cite{Lim:2016,Horsfield:2016,Holmstrom:2008,Holmstrom:2010} showed that also the ionization has a strong
dependence on crystal directions. Unfortunately these calculations are
too demanding computationally to obtain a full threshold map, and
hence we continue to work with the inference that the ionization
energy threshold correlates with the defect production threshold.

Consequently, the event rates obtained by
integrating the Radon transforms (\ref{eq:radontrafo}) and (\ref{eq:tradontrafo})
over energy become functions of the recoil direction. Contrary to Ref. \cite{Kadribasic:2017obi}, in the current work we did not
average the threshold energy surface over an angular interval.
Instead, we use the list of randomly sampled directions with the corresponding energy thresholds to compute the event rate $R=\int d\Omega (dR/d\Omega)$ directly as a Monte Carlo integral over the solid angle $\Omega$, as explained in more detail below.

\subsection{Purely velocity dependent interactions}

\begin{figure}
\begin{center}
\includegraphics[width=0.32\textwidth]{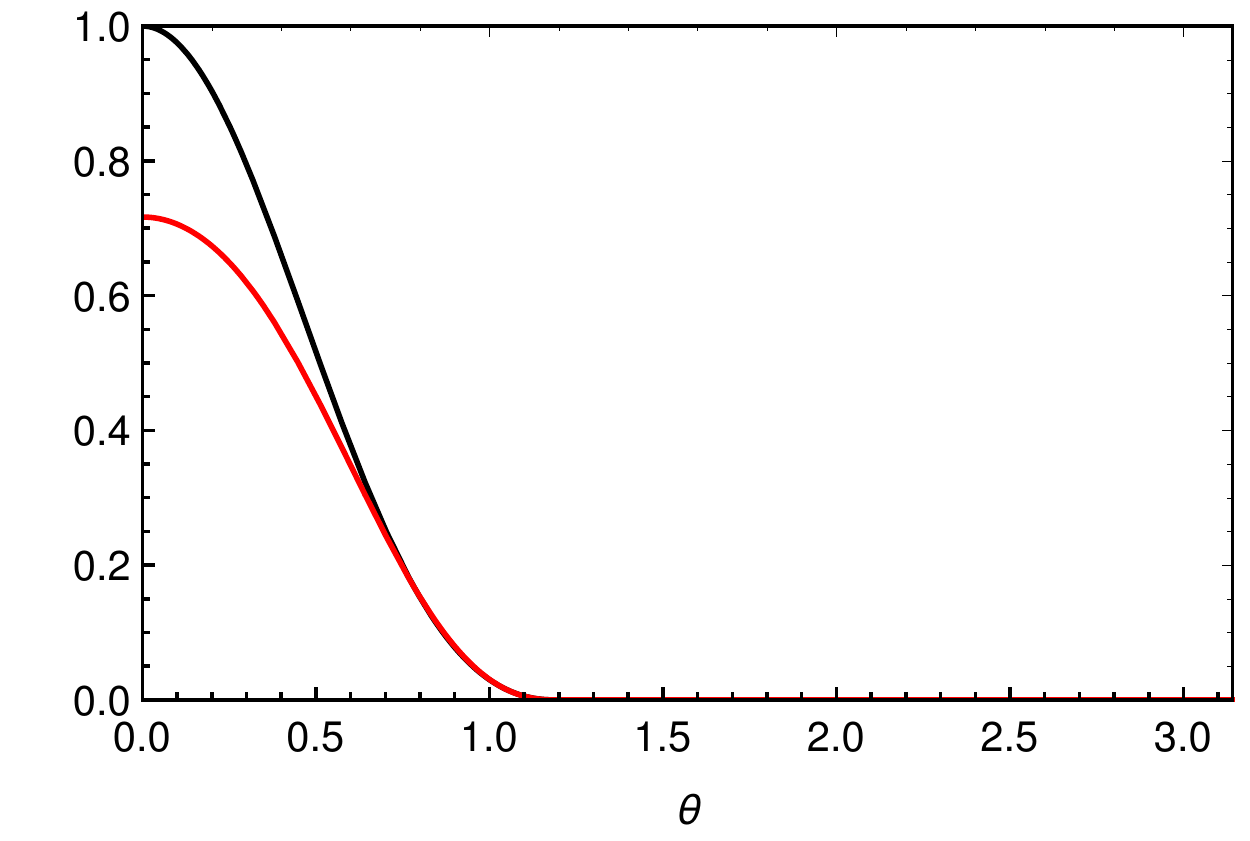}
\includegraphics[width=0.32\textwidth]{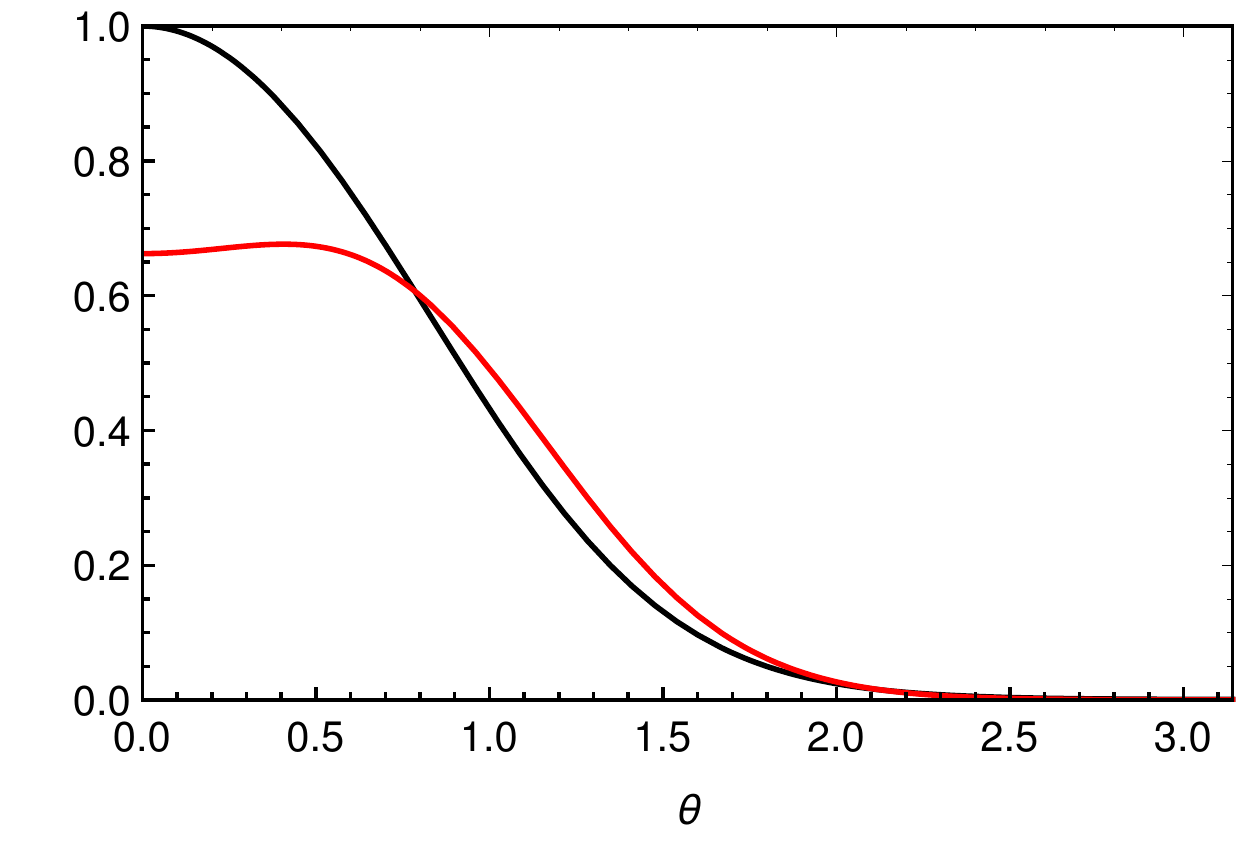}
\includegraphics[width=0.32\textwidth]{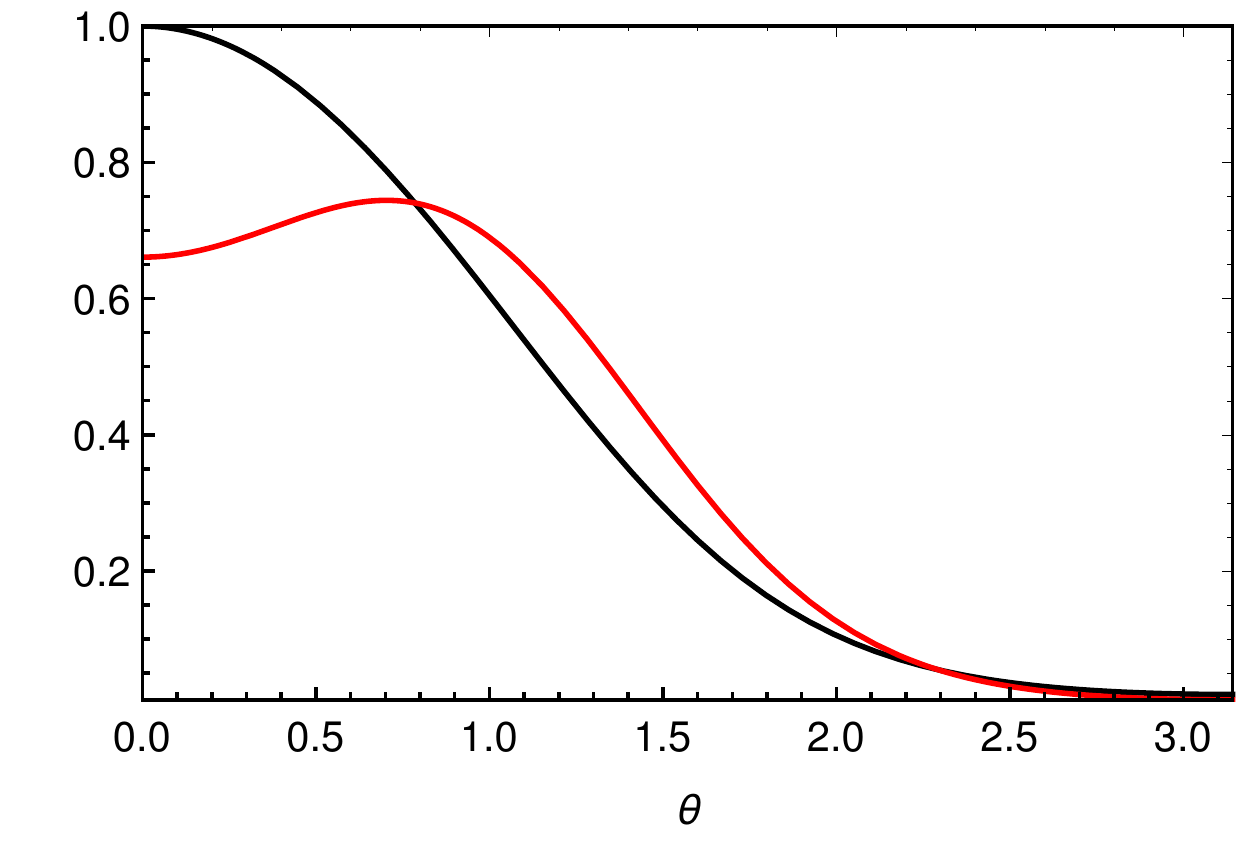}
\caption{Radon transform $\hat{f}_{\rm SHM}$ (black) and transverse Radon transform $\hat{f}^{\rm T}_{\rm SHM}$ (red), integrated from $E_{\rm min}=20$ eV to infinity, as a function of the angle $\theta$ between the average WIMP velocity and the recoil momentum, for $m_{DM}=0.4$ GeV (left), $m_{DM}=1$ GeV (center), $m_{DM}=5$ GeV (right). The atomic mass of Germanium, $m_{\rm N} = 72.64\, {\rm u} = 67.66\, {\rm GeV}$ has been assumed for the nuclear mass.}
\label{Radontransforms}
\end{center}
\end{figure}

To understand the general behavior of the angular differential rate as a function of the DM mass, we begin by showing the integrated Radon transforms (for $E_{\rm min}=20$ eV) of the velocity distribution $f_{\rm SHM}$, in figure \ref{Radontransforms} for various values of the WIMP mass. For the purpose of illustration, the functions have been arbitrarily normalized so that they match at the point $\theta = \pi/4$. We notice that for a small DM mass, both functions are strongly peaked towards forward recoil, $\theta = 0$, and hence the behavior of the angular differential rate in the low-mass region will be similar regardless of the $v_\perp^2$-dependence of the squared matrix element $|{\cal M}|^2$. For larger values of the DM mass both distributions become broader, and the transverse Radon transform develops a maximum at some non-zero recoil angle.

\begin{figure}
\begin{center}
\includegraphics[width = 0.45\textwidth]{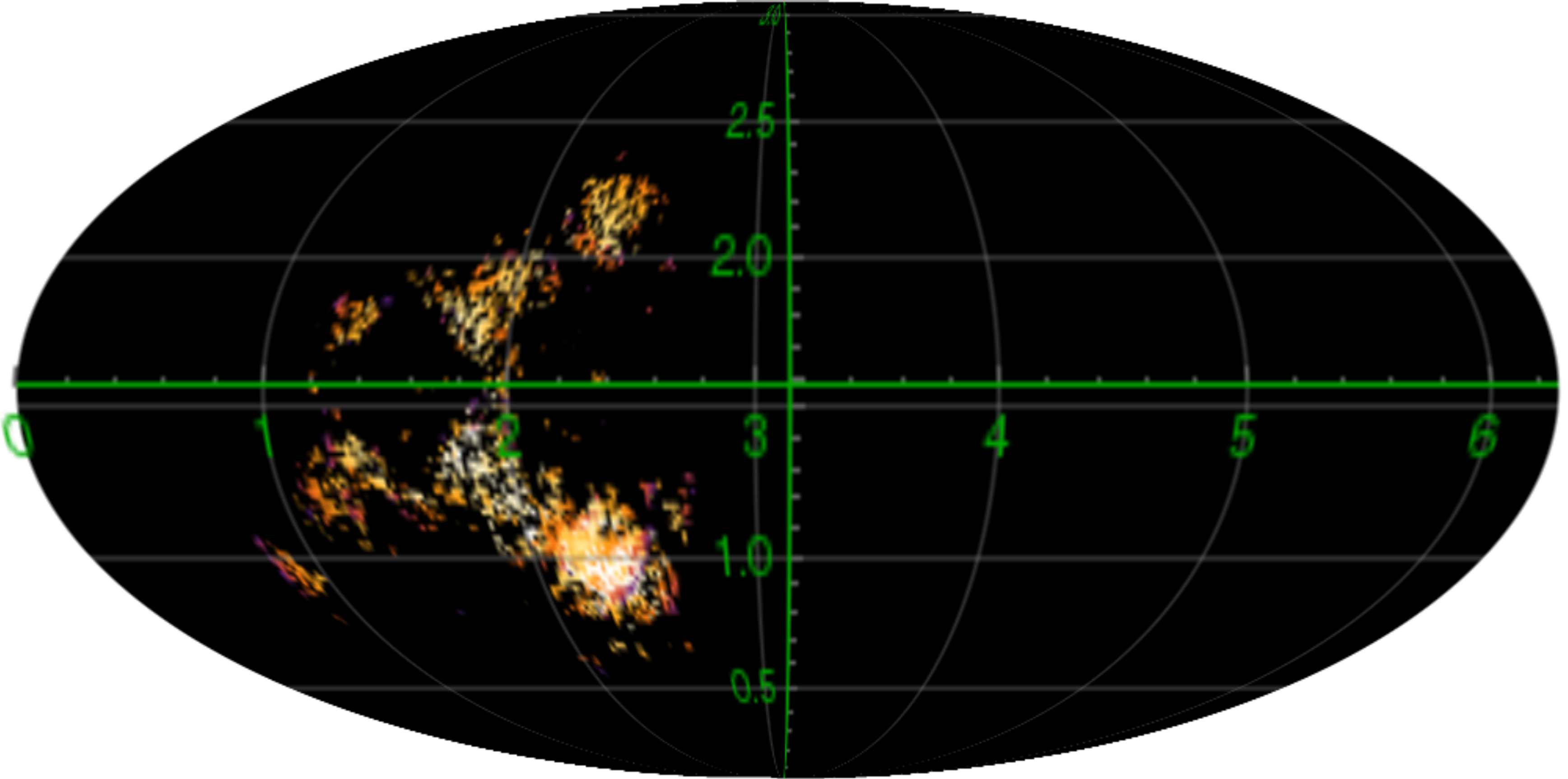}
\includegraphics[width = 0.45\textwidth]{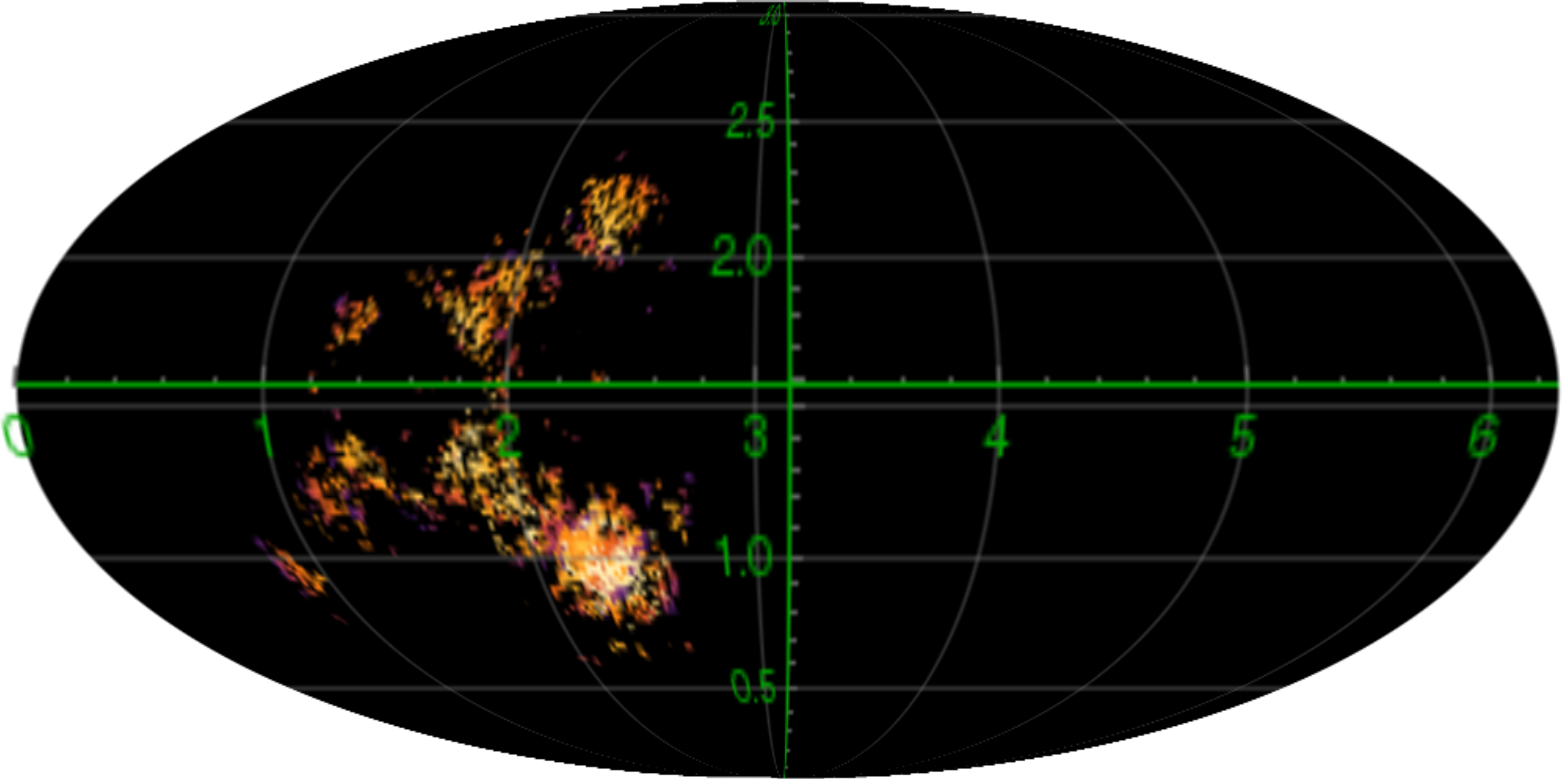}
\includegraphics[width = 0.046\textwidth]{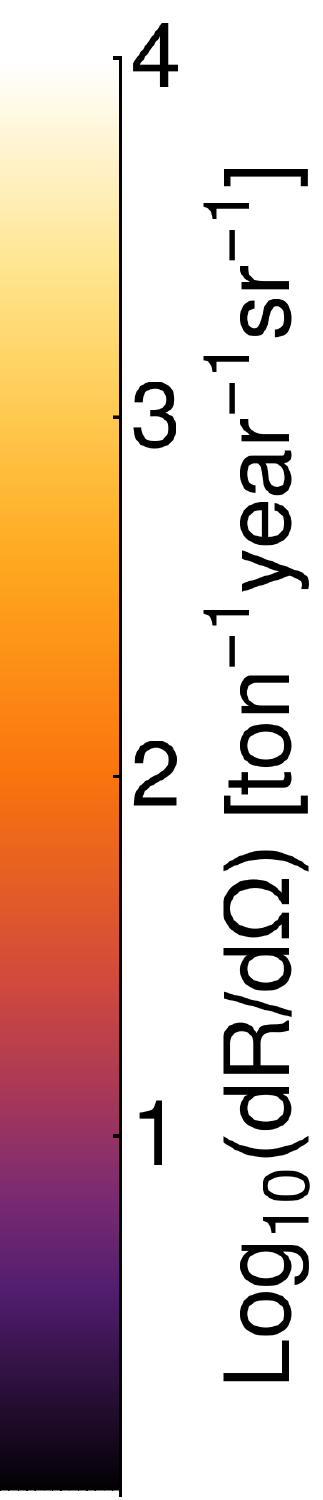} \\
\includegraphics[width = 0.45\textwidth]{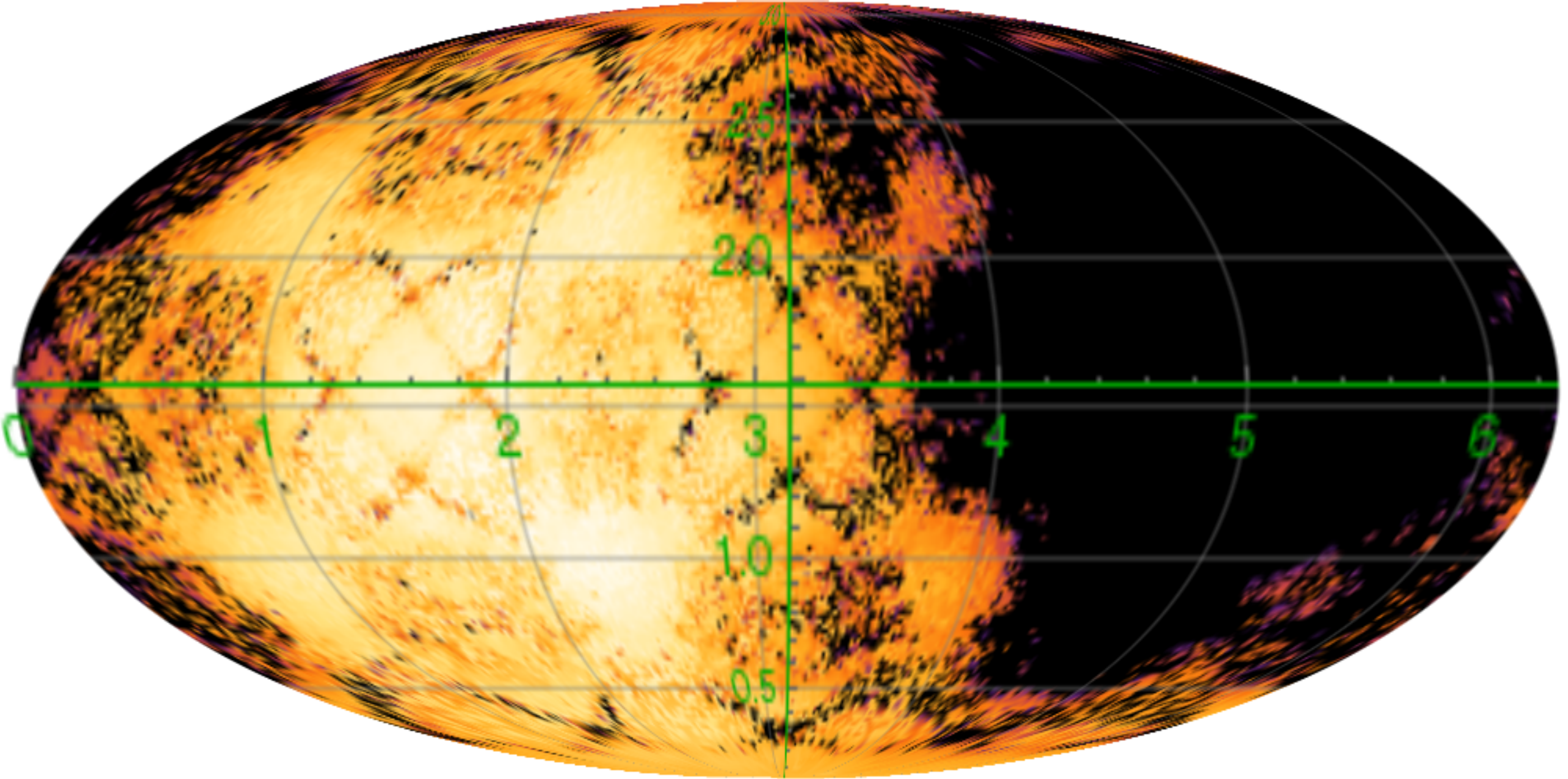}
\includegraphics[width = 0.45\textwidth]{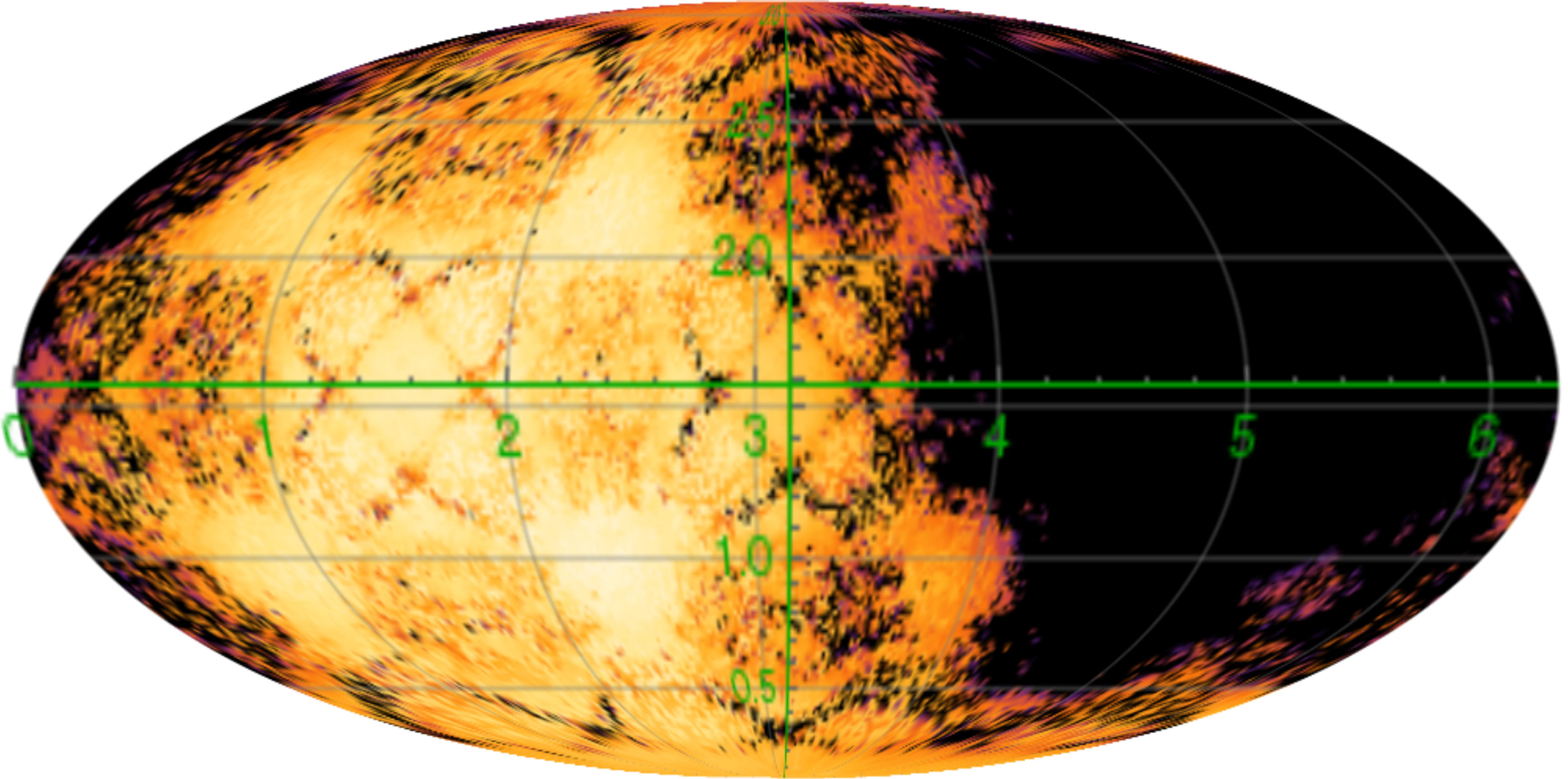}
\includegraphics[width = 0.046\textwidth]{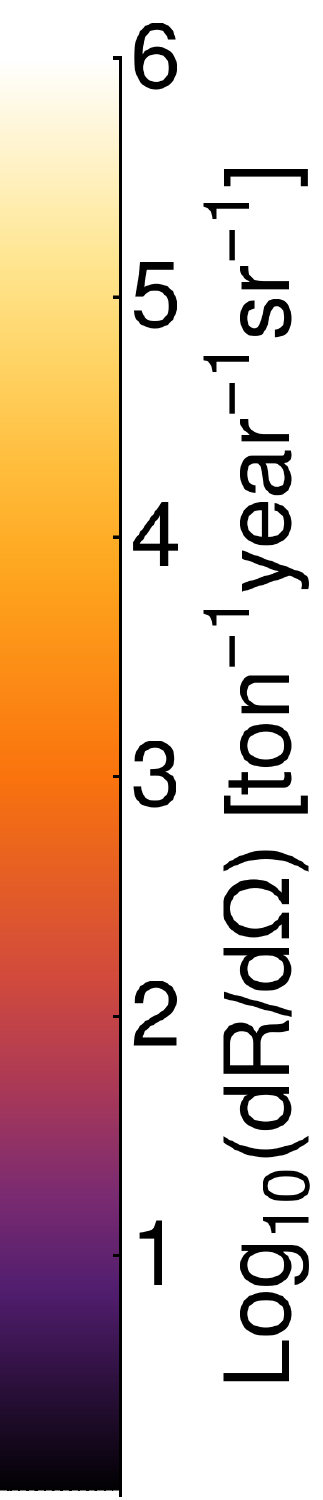} \\
\mbox{\hspace{0.2cm} \includegraphics[width = 0.45\textwidth]{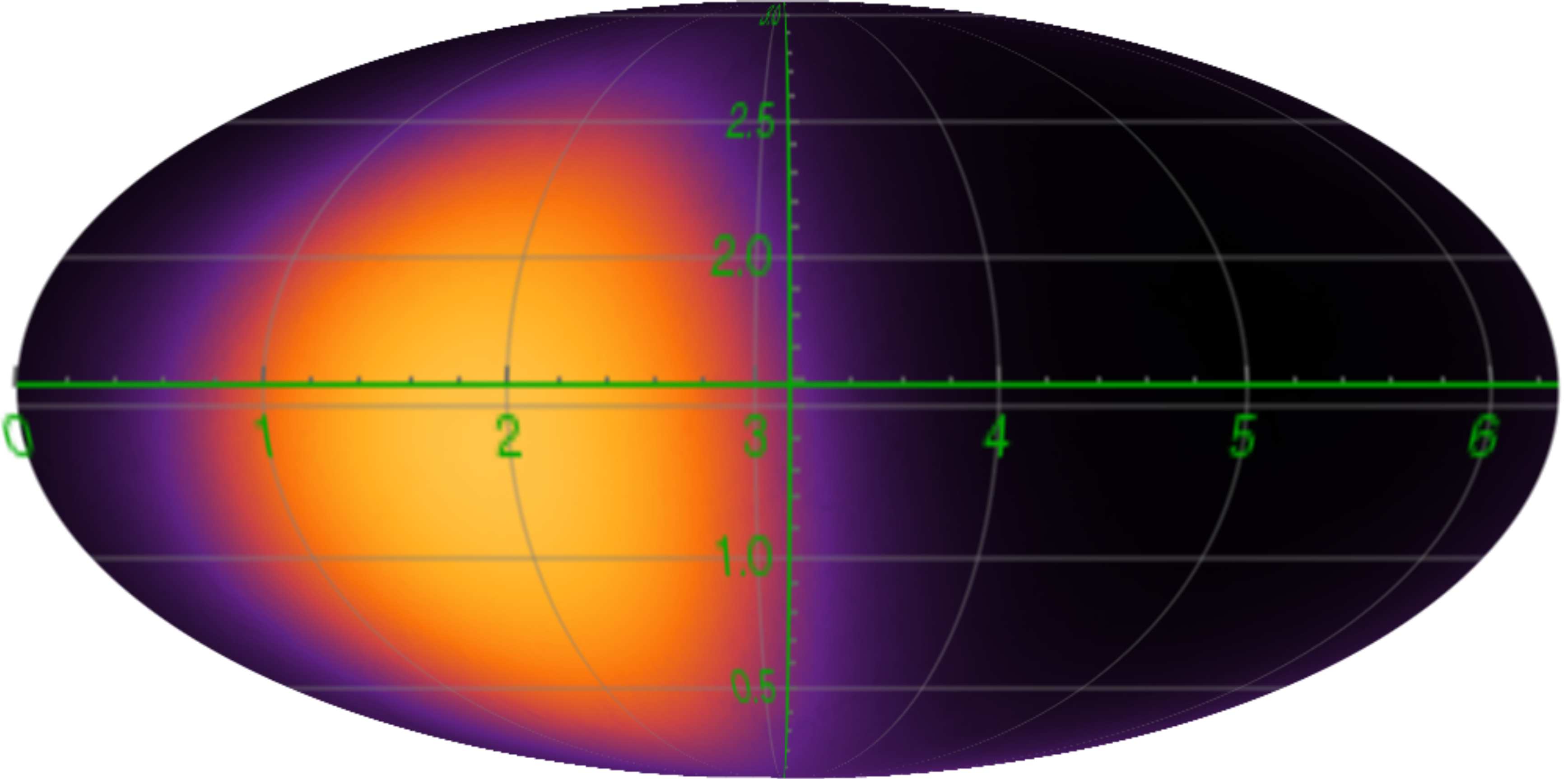}
\includegraphics[width = 0.45\textwidth]{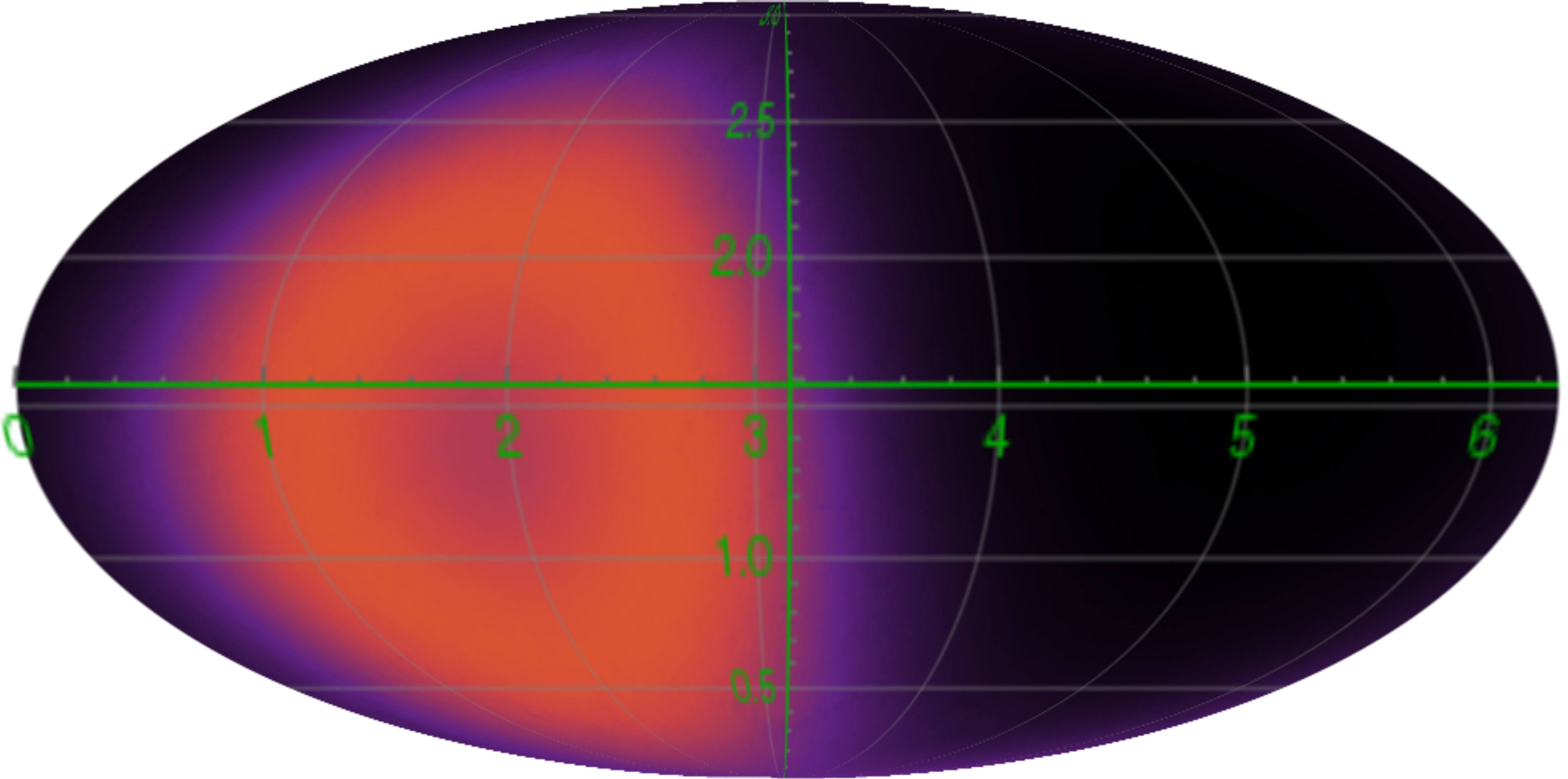}
\includegraphics[width = 0.07\textwidth]{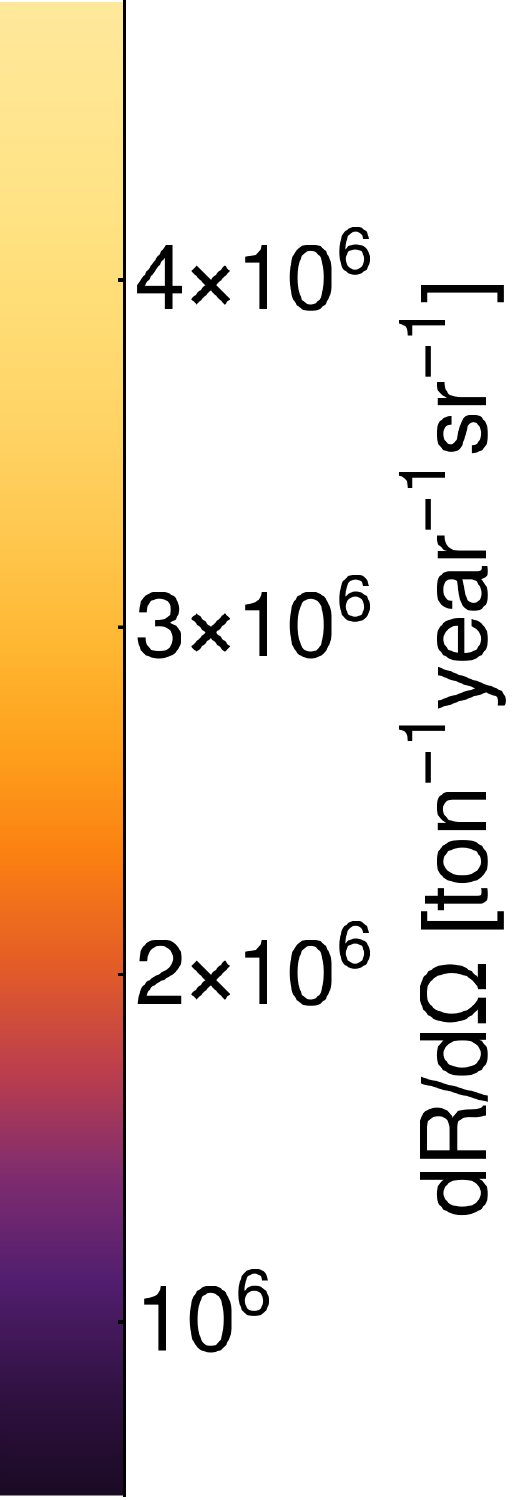}} \\
\caption{Angular differential event rate for the $1$-operator (left) and the $v_\perp^2$-operator  (right)
for $m_{DM}=0.3$ GeV (top row), $m_{DM}=0.5$ GeV (middle row) and
$m_{DM}=5$ GeV (bottom row).}
\label{fig:dir rates}
\end{center}
\end{figure}

Then, to demonstrate the effect of the direction-dependent energy threshold, we show the angular event rate for the $v^0$- (the Radon transform) and $v_\perp^2$- (the transverse Radon transform) interactions in figure~\ref{fig:dir rates} for various values of the WIMP mass. These figures are obtained by
integrating the Radon transforms over energy, with $E_{\rm min}$ in each direction given by the data shown in figure \ref{fig:GE threshold}. For these calculations, we have assumed the SHM velocity distribution and a Germanium detector on the SNOLAB site $(46.4719^\circ {\rm N}, 81.1868^\circ {\rm W})$ on September 6, 2015 at 18:00. The event rates correspond to the Spin-independent DM-nucleon cross section $a_1\sigma_0 = 10^{-39}\ {\rm cm}^2$ in the $v^0$-case, and $b_1\sigma_0 = 10^{-33}\ {\rm cm}^2$ for the $v_\perp^2$-interaction.

We notice that the distributions for the small DM masses, shown in the top row of the figure, are basically indistinguishable by eye, and centered towards the average direction of the DM wind. As the DM mass is increased, the angular recoil distribution becomes wider. Eventually, for large enough $m_{\rm DM}$, the information of the directionality of the energy threshold becomes practically undetectable, as is evident from the figures on the bottom row. Towards large $m_{\rm DM}$, the off-zero maximum of the transverse Radon transform manifests as the ring-like feature around the direction of the DM wind, visible in the bottom right figure.

\begin{figure}
\begin{center}
\includegraphics[width=0.45\textwidth]{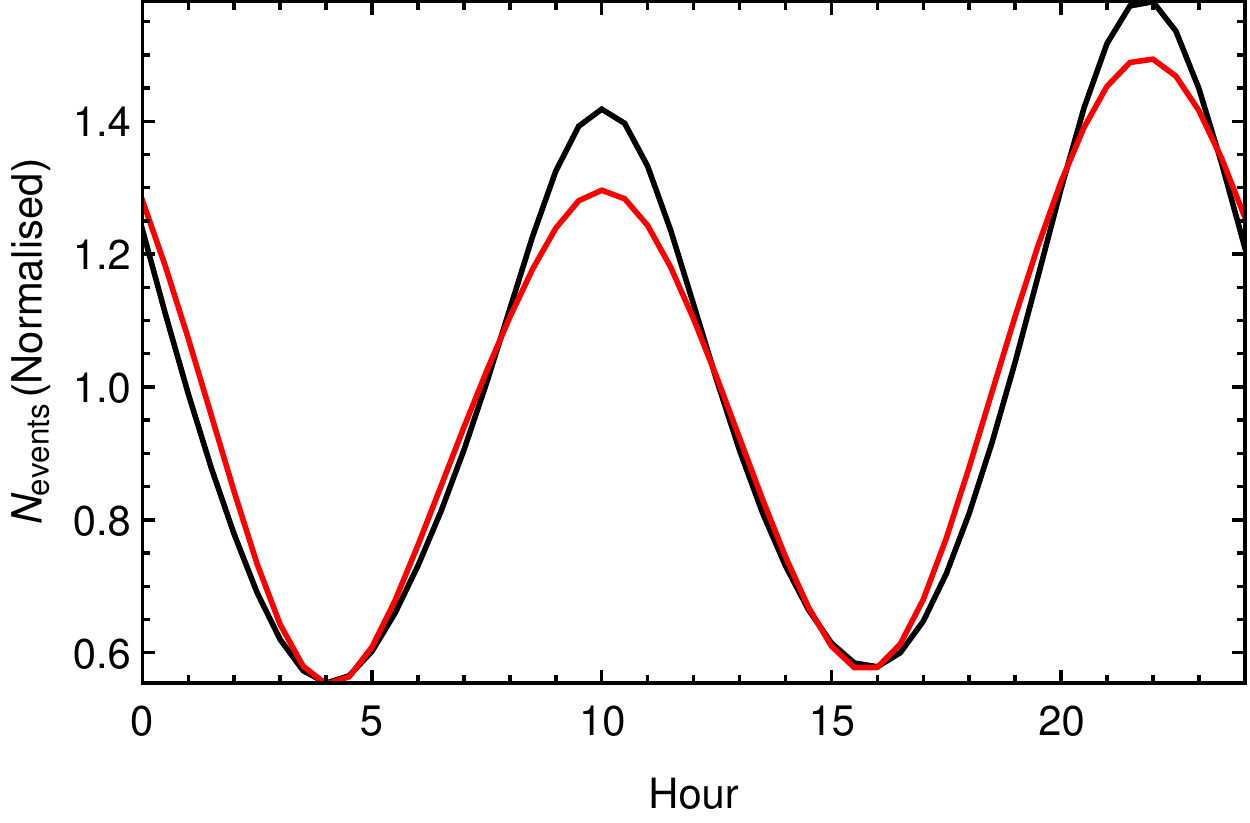}
\includegraphics[width=0.45\textwidth]{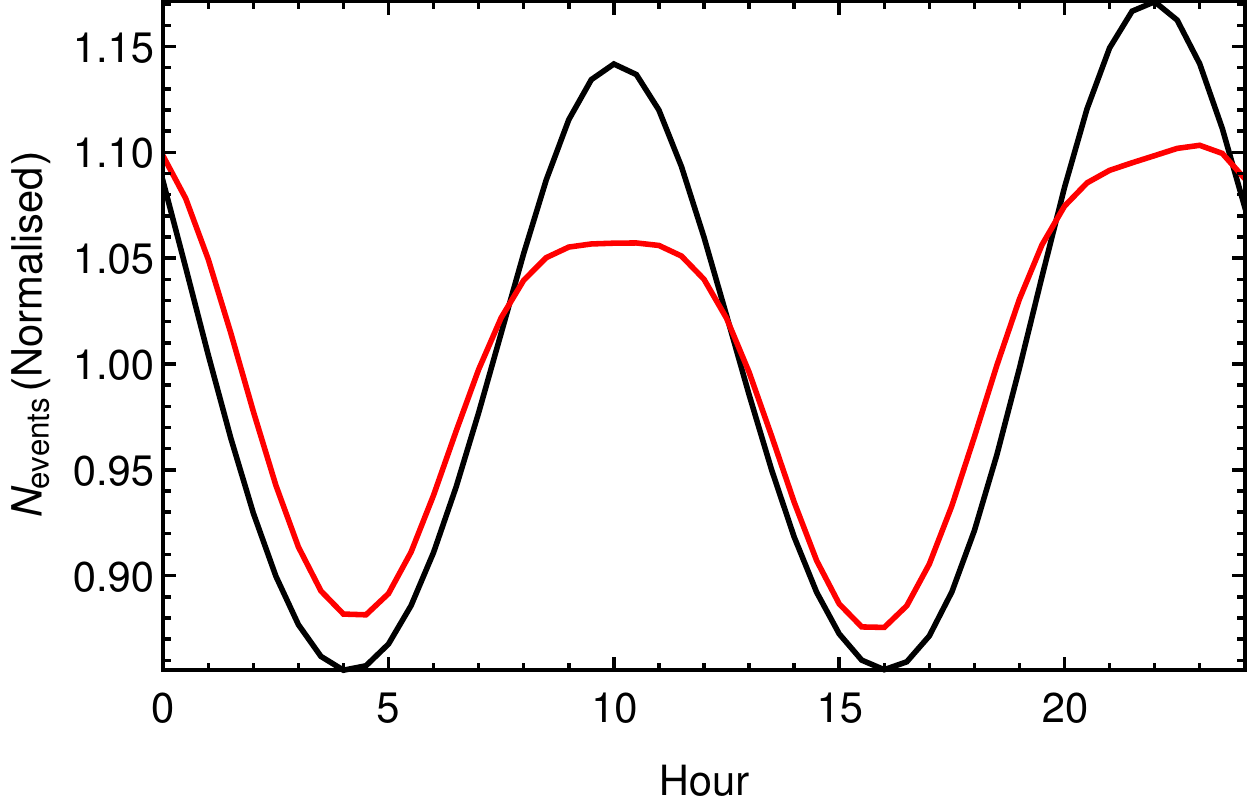}\\
\includegraphics[width=0.45\textwidth]{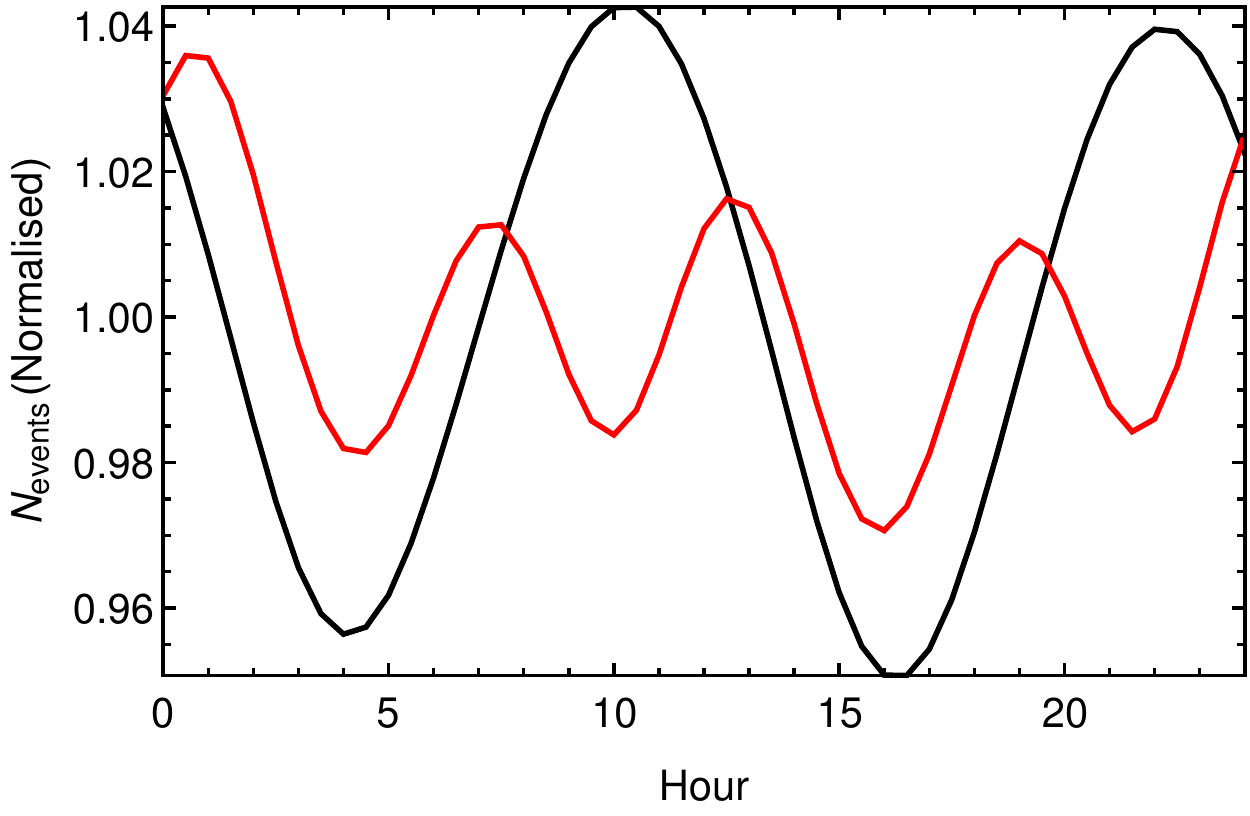}
\includegraphics[width=0.45\textwidth]{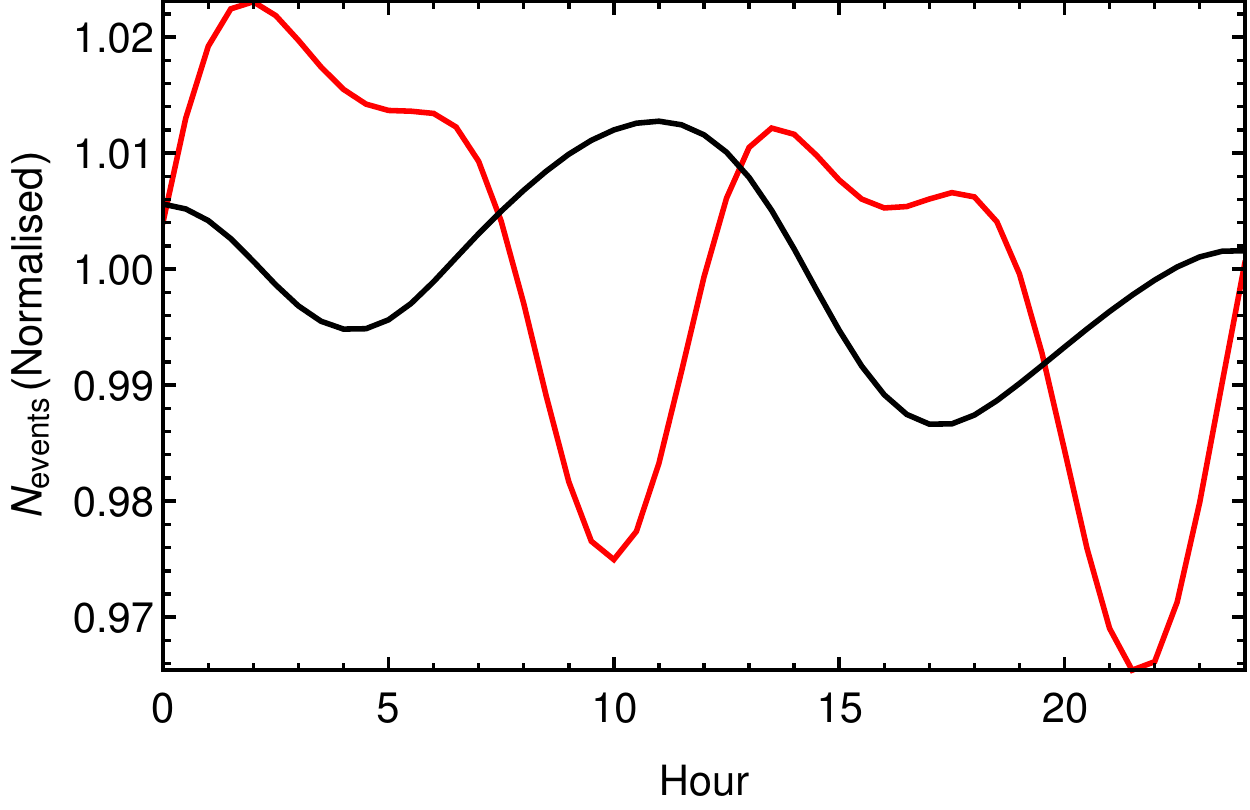}\\
\includegraphics[width=0.45\textwidth]{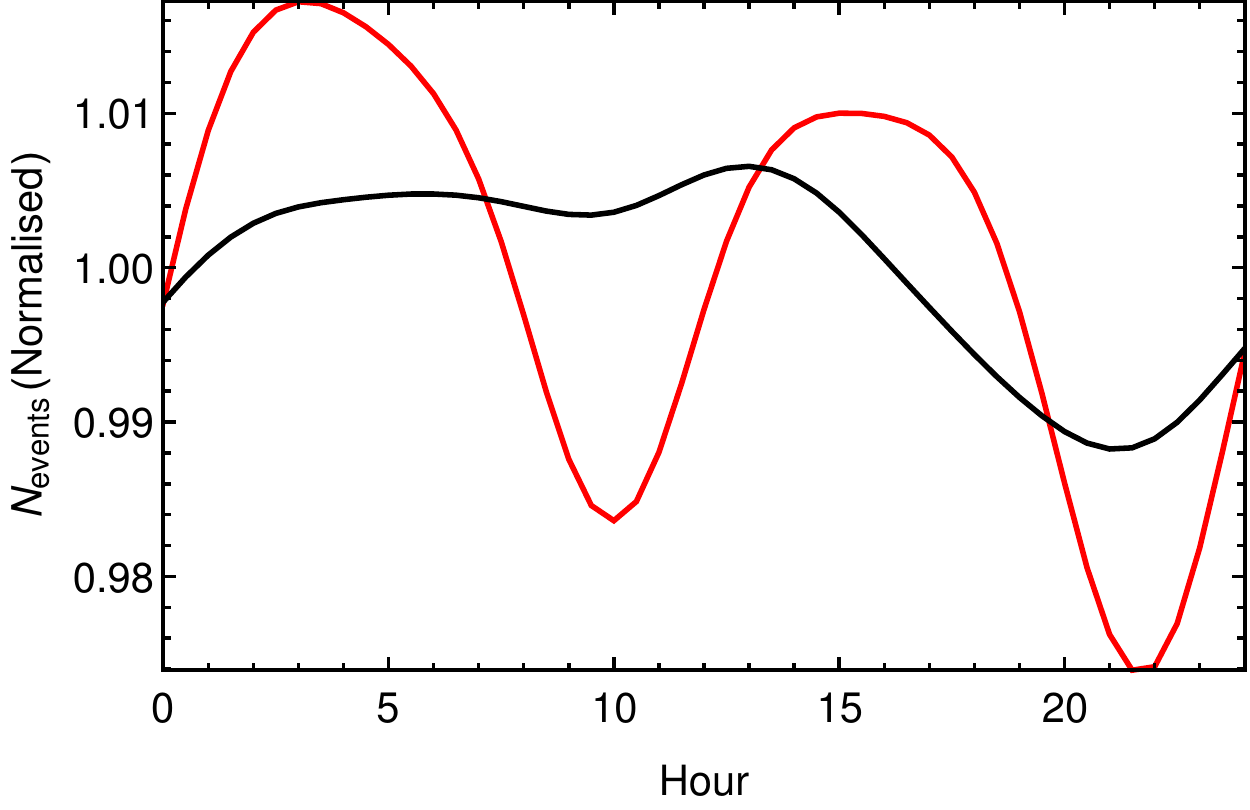}
\includegraphics[width=0.45\textwidth]{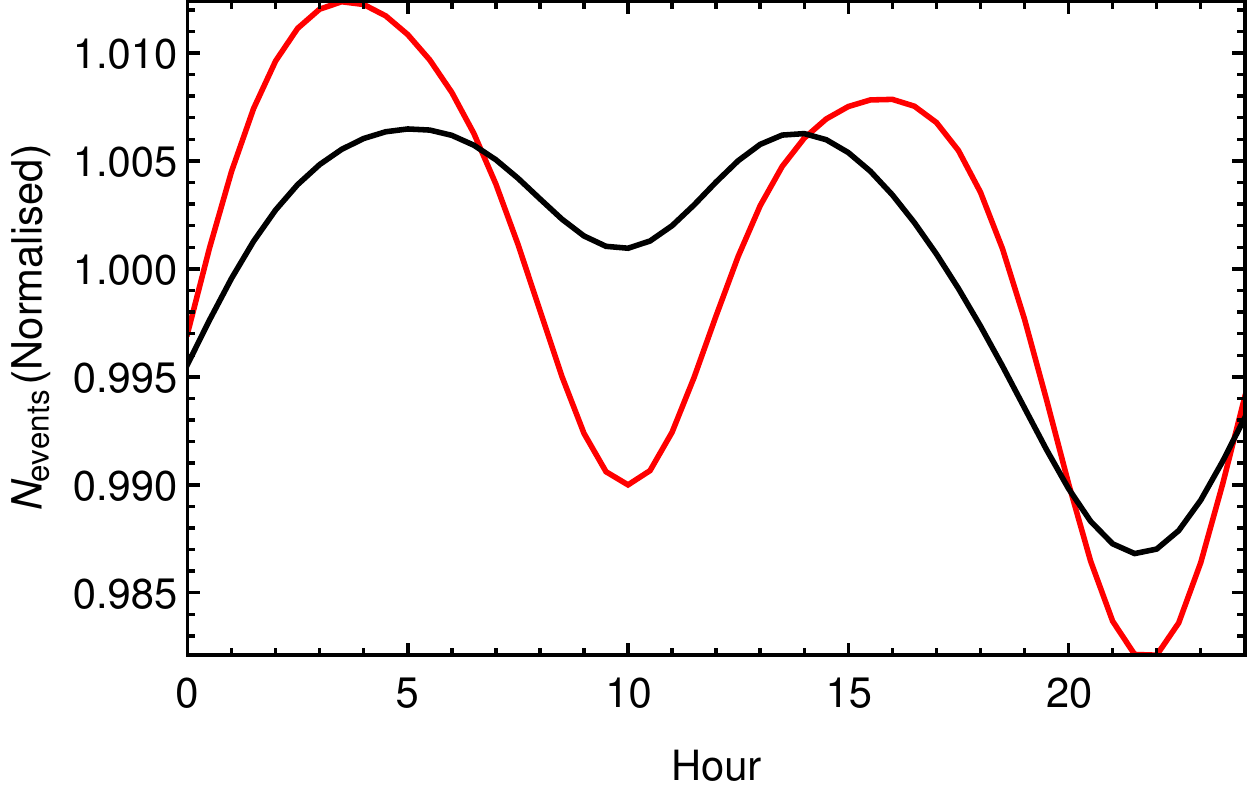}

\caption{Normalised event rate $R(t)/\langle R \rangle$ for $v^0$ (black) and $v_\perp^2$ (red) interactions as
a function of time for $m_{DM}=0.3$ GeV (top left), $m_{DM}=0.33$ GeV (top right), $m_{DM}=0.36$ GeV (middle left), $m_{DM}=0.4$ GeV (middle right), $m_{DM}=0.45$ GeV (bottom left), $m_{DM}=0.5$ GeV (bottom right).}
\label{fig:daily}
\end{center}
\end{figure}

As the Earth rotates around its axis, the direction of the DM wind with respect to the lab-frame modulates. Due to the directional dependence of the event rate shown above, this results in a diurnal modulation of the integrated event rate. Figure \ref{fig:daily} shows the diurnal modulation in the event rate for various values of the WIMP mass during September 6, 2015. We compute the event rate $R=\int d\Omega (dR/d\Omega)$ by a Monte Carlo integral over the solid angle $\Omega$, utilizing the list of 84936 randomly sampled directions with the corresponding energy thresholds. Energy integrals of the Radon transforms are evaluated for each sampled point $(\theta_i,\phi_i)$ on the surface of the unit-sphere, with the corresponding value for the threshold energy $E_{\rm min}(\theta_i,\phi_i)$ obtained from the list. For each $(\theta_i,\phi_i)$-point this procedure yields the corresponding differential event rate $dR(\theta_i,\phi_i)$. The total event rate is then obtained as the sum over the points in the list:
\be
R(t) = \frac{4\pi}{N_{\rm points}}\sum\limits_{i=1}^{N_{\rm points}} dR(\theta_i,\phi_i,t),
\ee
where the dependence on time $t$ follows from the time-dependence of the laboratory's motion in the galactic rest frame $\bvec{V}(t)$, as explained in the appendix \ref{app:radontrans}.
We have checked that the number of points is sufficient for an accurate integral: Already for $N_{\rm points}=5000$ the result of the sum is within 3\% of the result for using the total $\sim 85 000$ points in the list.

As expected from the discussion above, the shape of the diurnal modulation signal for the smallest DM mass displayed here, $m_{\rm DM} = 0.3$ GeV in the top left figure, is very similar for the $v^0$- and $v_\perp^2$-interactions. As the DM mass is increased, the expected modulations become more different, but the amplitude of the signal quickly drops below $\mathcal{O}(1\%)$, and thus undetectable. However, for the DM mass, $340\  {\rm MeV}\lesssim m_{\rm DM} \lesssim 450\ {\rm MeV}$, the shape of the daily modulation signal can be used to determine the velocity-dependence of the scattering amplitude.

\begin{figure}
\begin{center}
\includegraphics[width = 0.32\textwidth]{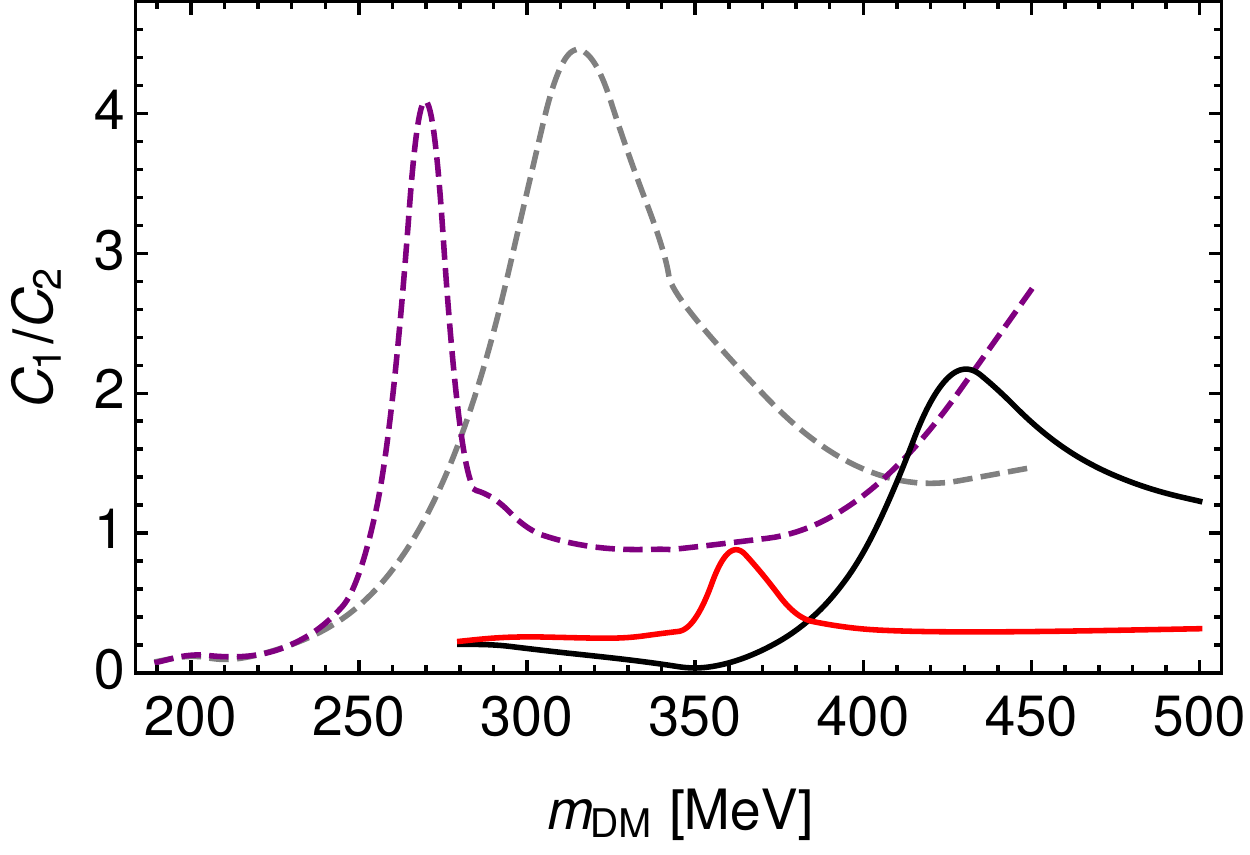}
\includegraphics[width = 0.32\textwidth]{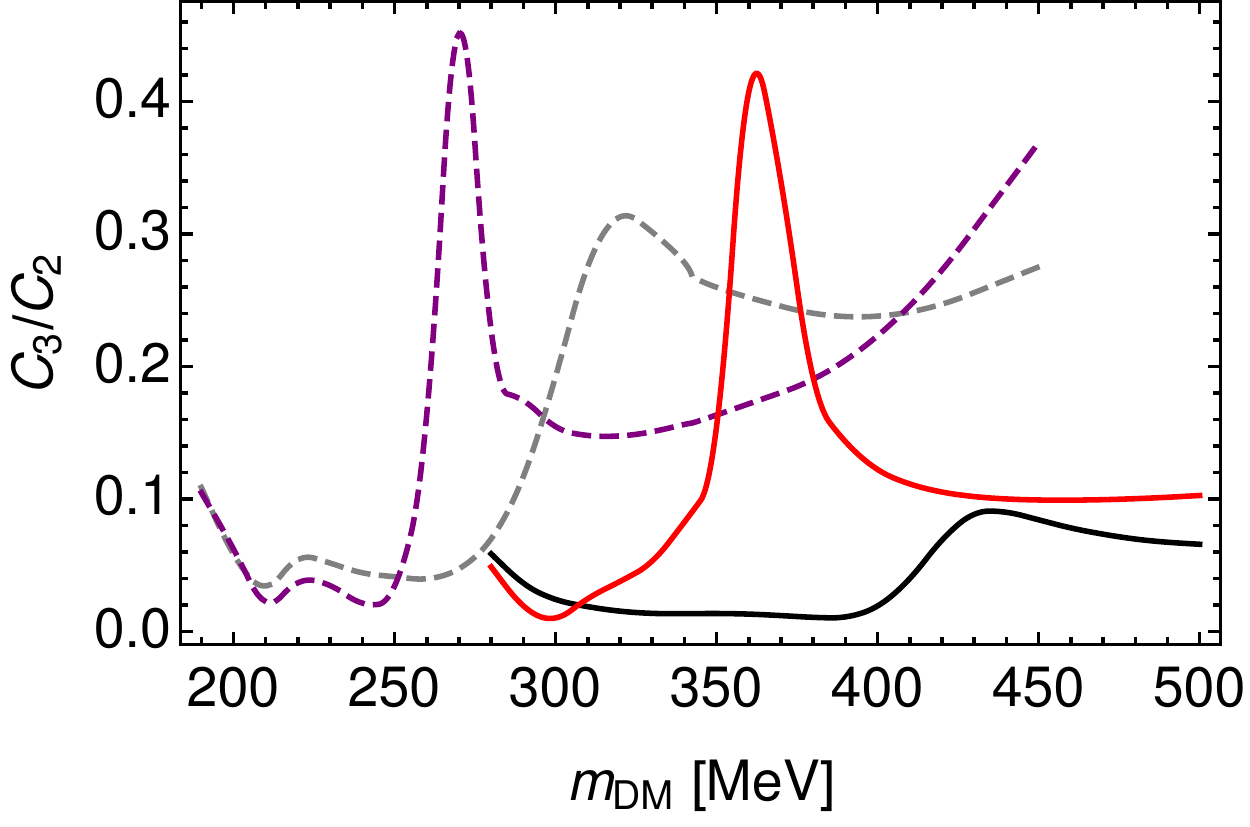}
\includegraphics[width = 0.32\textwidth]{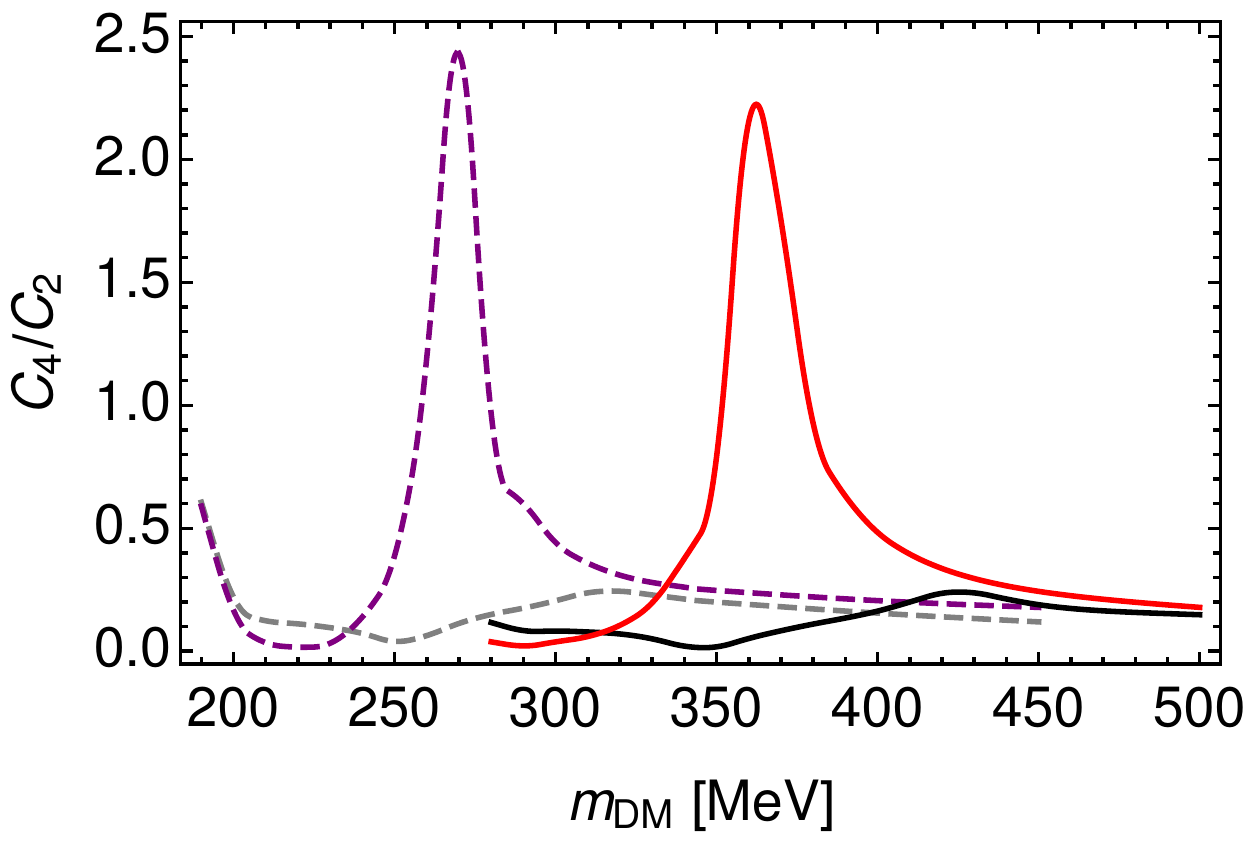}
\caption{The ratios of the Fourier-components $C_1/C_2$ (left), $C_3/C_2$ (center) and $C_4/C_2$ (right), for the $v^0$-interaction  (black line) and $v_\perp^2$-interaction (red line), as a function of the DM mass. The gray and purple dashed lines show the same ratios for a Silicon detector.}
\label{fig:fouriermodes}
\end{center}
\end{figure}

To assess the feasibility of velocity-dependence detection from the shape of the daily modulation signal, we analyze the Fourier-components of the daily event rates, $C_n = \sqrt{a_n^2+b_n^2}$, where
\be
a_n = \int\limits_{0}^{1}R(t)\cos\left(2\pi n t \right) dt, \quad
b_n = \int\limits_{0}^{1}R(t)\sin\left(2\pi n t \right) dt,
\ee
where $t$ is time in units of day. We show the ratios $C_1/C_2$, $C_3/C_2$ and $C_4/C_2$ in figure \ref{fig:fouriermodes}, for the $v^0$-interaction in black and $v_\perp^2$-interaction in red, as a function of the DM mass. For any value of the DM mass above $m_{\rm DM}\gtrsim 340\ {\rm MeV}$ at least one of the ratios is substantially different to allow separation of the $v^0$ and $v_\perp^2$ interactions, as long as the Fourier-components can be reliably reconstructed from the data. Above $m_{\rm DM}\gtrsim 450\ {\rm MeV}$ the amplitude of the daily modulation rate drops below 1\%, and the reconstruction of the Fourier-components becomes prohibiting in the required scale of the experiment. Figure \ref{fig:fouriermodes} also shows the ratios of the Fourier components for a Silicon detector, where the $v^0$-interaction is shown by the gray dashed line, and the $v_\perp^2$-interaction by the purple dashed line. Due to the smaller atomic mass of Silicon compared to Germanium, the Si sensitive region falls at the lower values of the DM mass. We identify the range of $250\  {\rm MeV}\lesssim m_{\rm DM} \lesssim 350\ {\rm MeV}$ as the region where the velocity-dependence of the operator can be identified in Silicon.

\begin{figure}
\begin{center}
\includegraphics[width=0.45\textwidth]{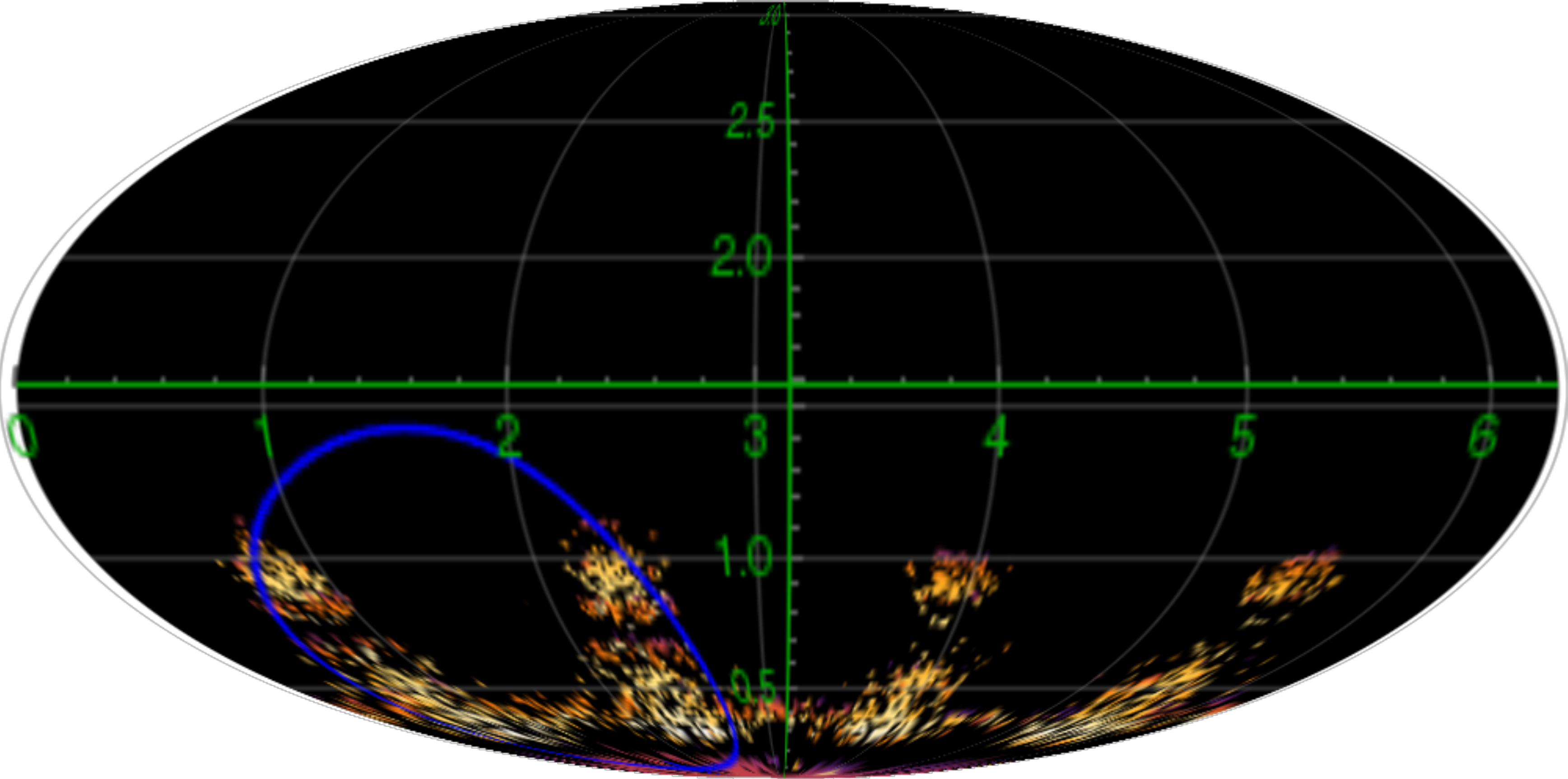}
\includegraphics[width=0.45\textwidth]{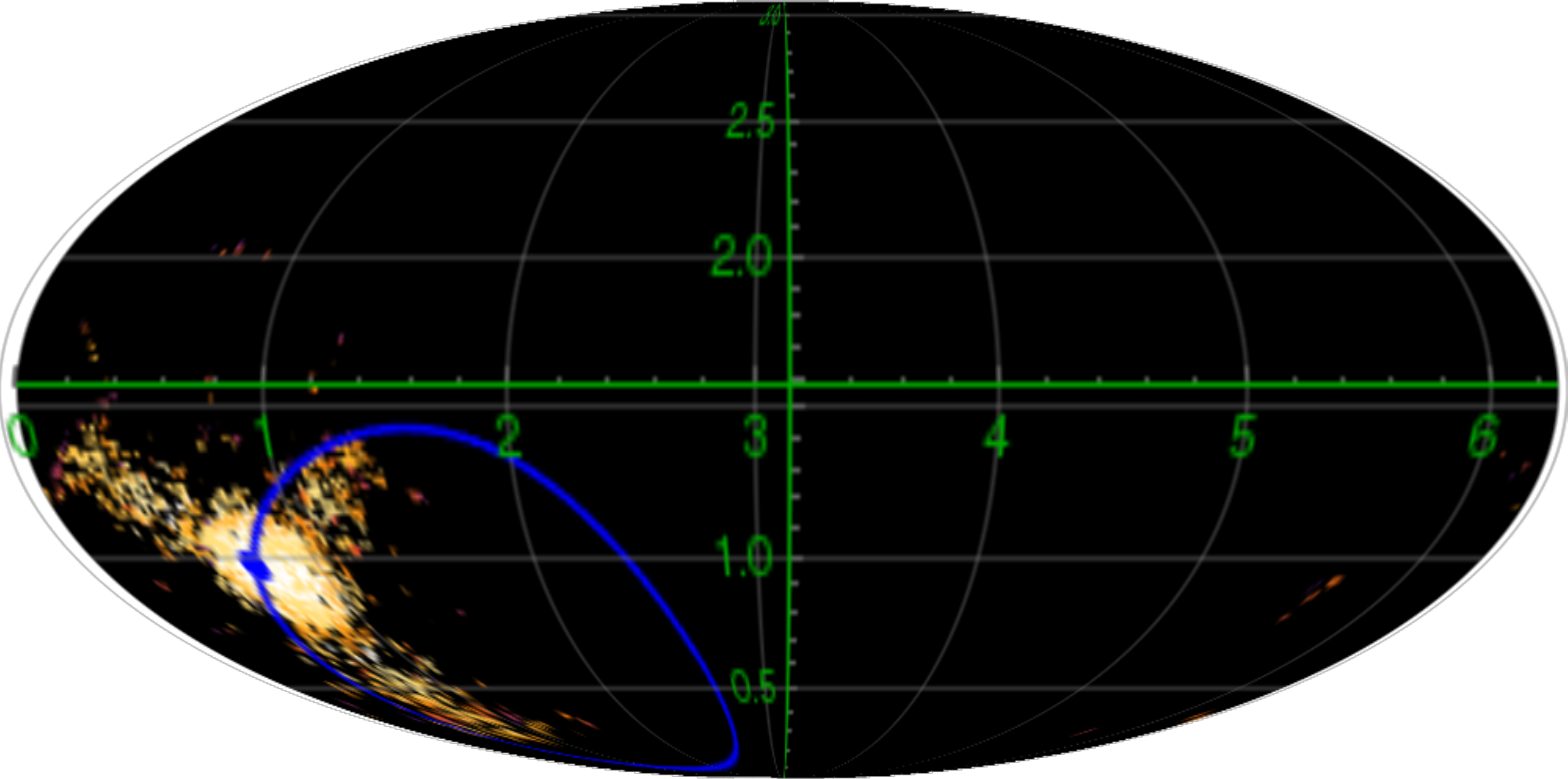}\\
\includegraphics[width=0.45\textwidth]{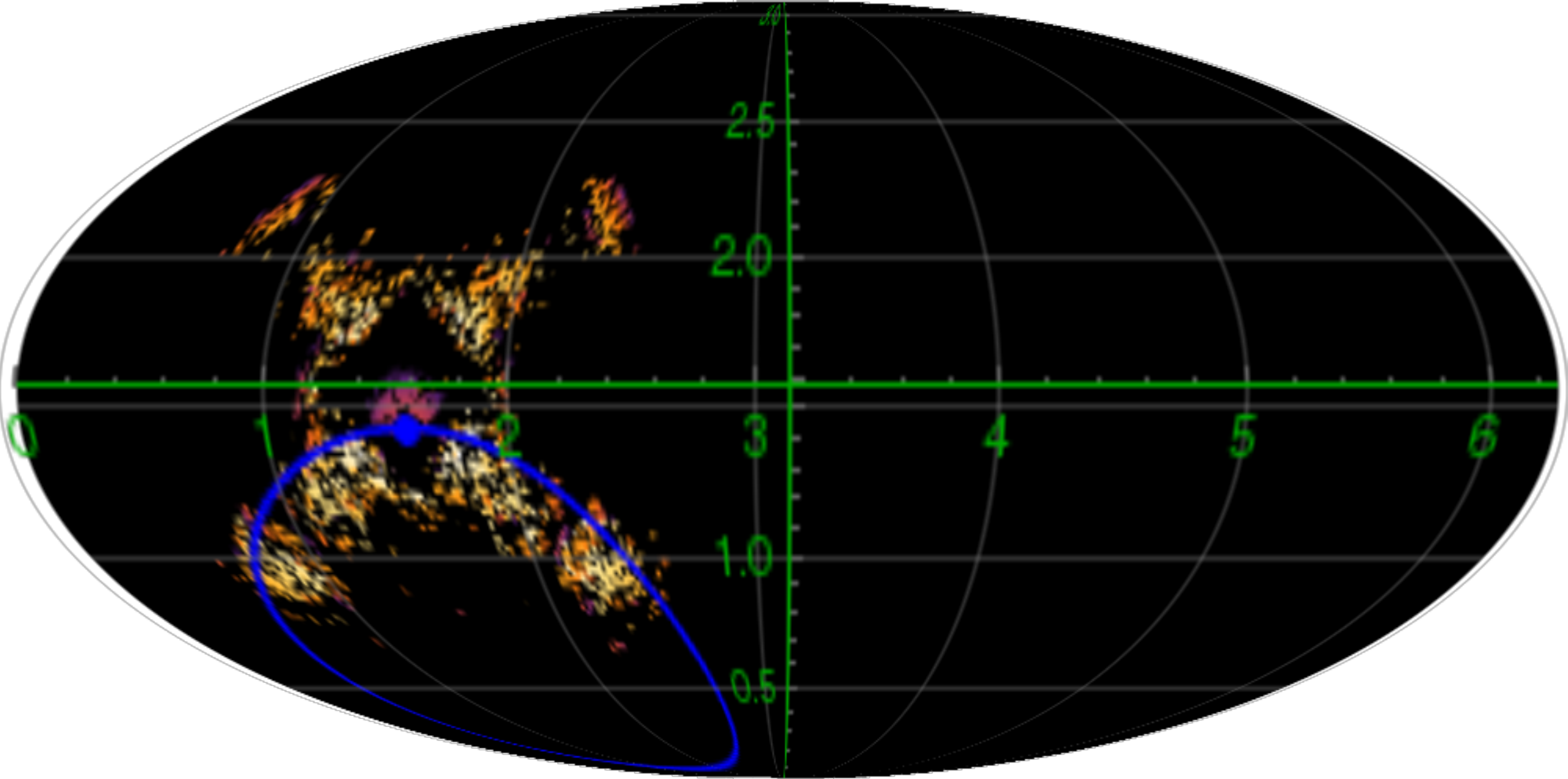}
\includegraphics[width=0.45\textwidth]{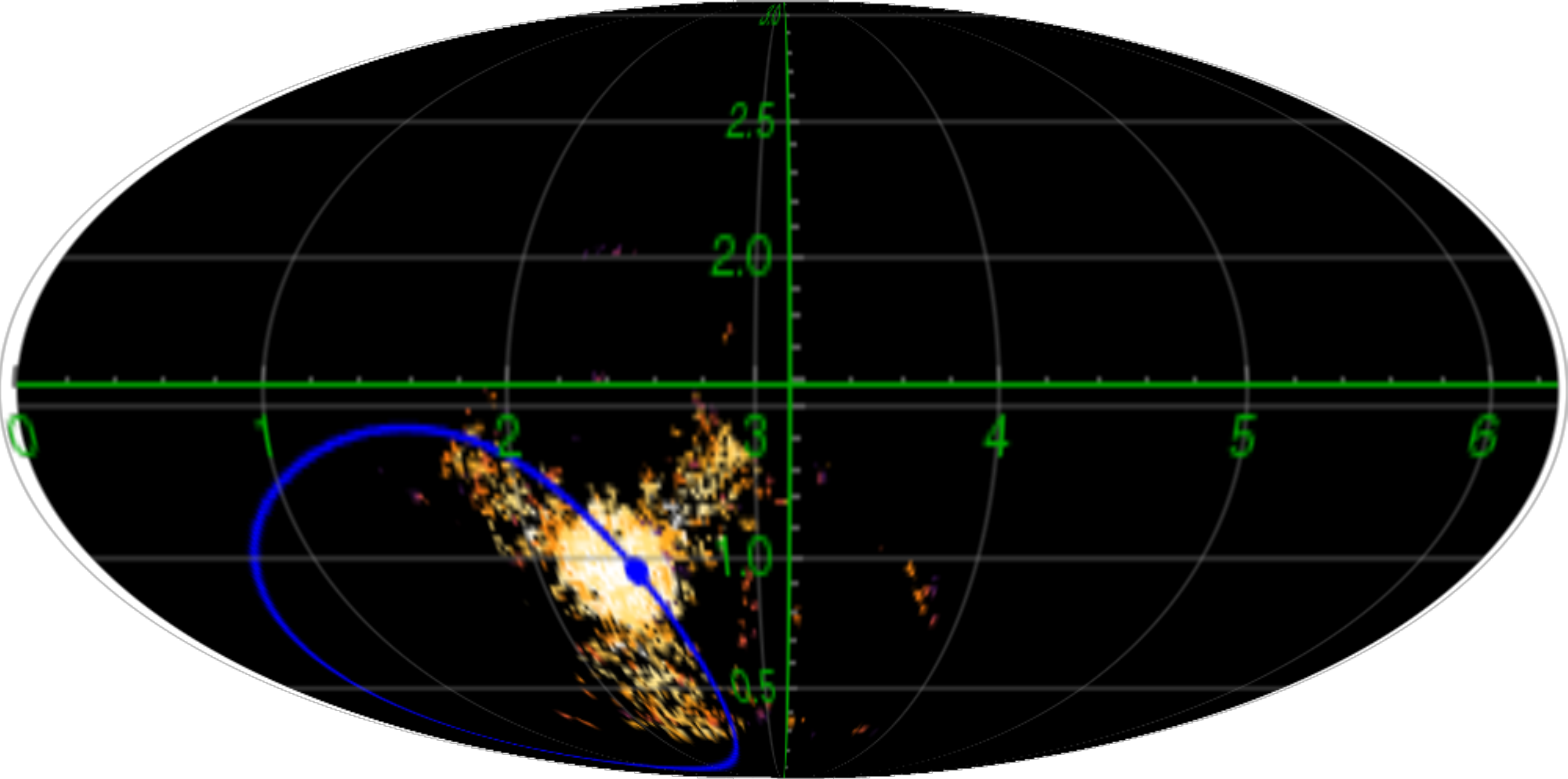}\\
\includegraphics[width = 0.33\textwidth]{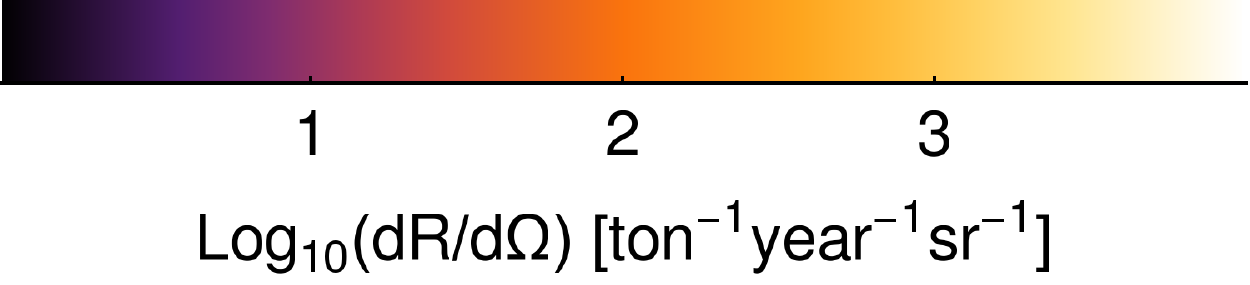}
\caption{Differential event rate as a function of direction for the $v^0$-interaction for $m_{DM}=0.3$ GeV at the times of minimum and maximum total event rate: 04:00h (top left), 10:00h (top right), 16:00h (bottom left), 22:00h (bottom right). The blue dot shows the average direction of the incoming DM particles in the lab-frame. During the day the DM direction covers the curve shown in blue.}
\label{fig:daily minmax}
\end{center}
\end{figure}

To understand how the shape of the daily modulation signal arises, we show in figure \ref{fig:daily minmax} the event rate for the $v^0$-interaction as a function of the recoil direction, for $m_{DM}=0.3$ GeV, corresponding to the top left panel of figure \ref{fig:daily}, at the moments of minimum and maximum event rates (at 04:00, 10:00, 16:00, 22:00 hours), assuming $a_1\sigma_0 = 10^{-39}\ {\rm cm}^2$. Comparing to figure \ref{fig:GE threshold}, we see that the maximum event rates (corresponding to 10:00 and 22:00), shown on the right column of figure \ref{fig:daily minmax}, occur when the direction of the DM wind, shown by the blue dot in figure \ref{fig:daily minmax}, coincides with the low threshold energy directions that appear as the dark spots in figure \ref{fig:GE threshold}. Respectively, the minima of the event rate (at 4:00 and 16:00, shown on the left) occur when the direction of the DM wind is maximally far away from the low threshold regions. The blue curve in the figure shows the path of the direction of the DM wind on the unit sphere during the 24 hour period.

\begin{figure}
\begin{center}
\includegraphics[width=0.49\textwidth]{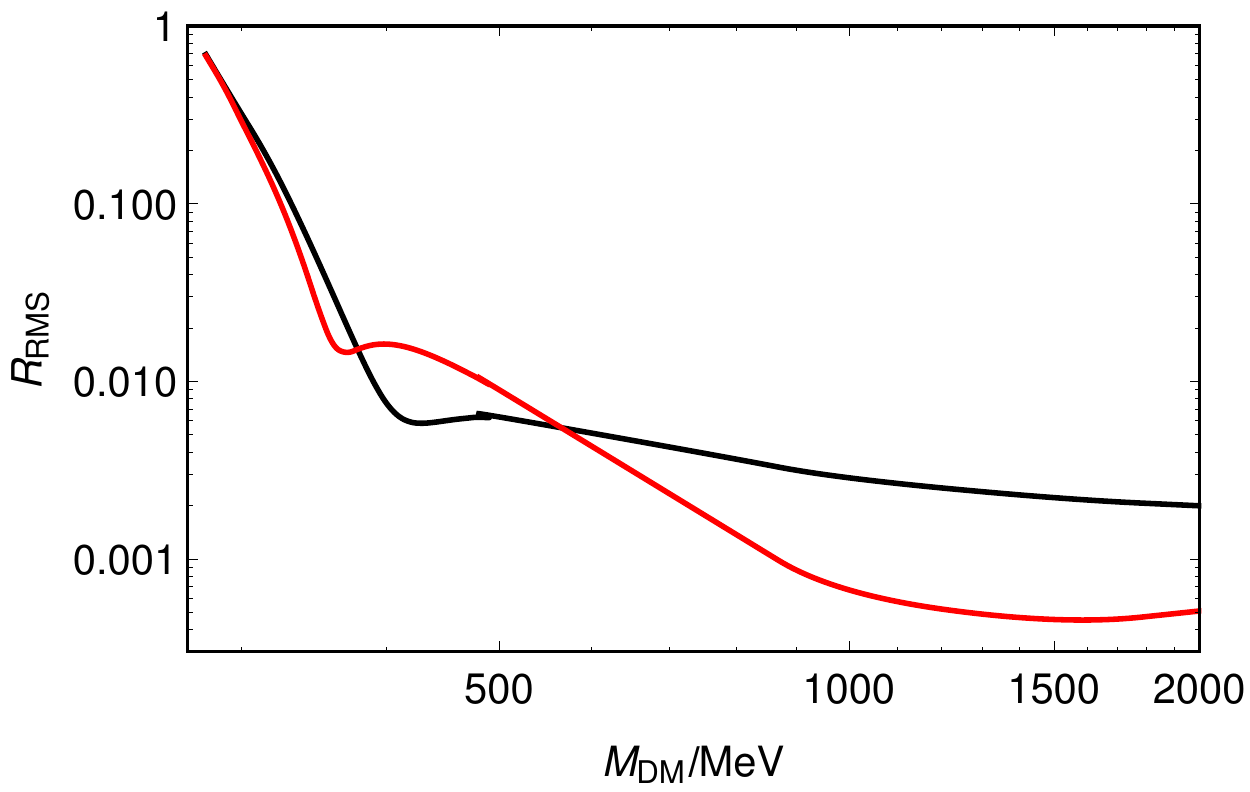}
\includegraphics[width=0.49\textwidth]{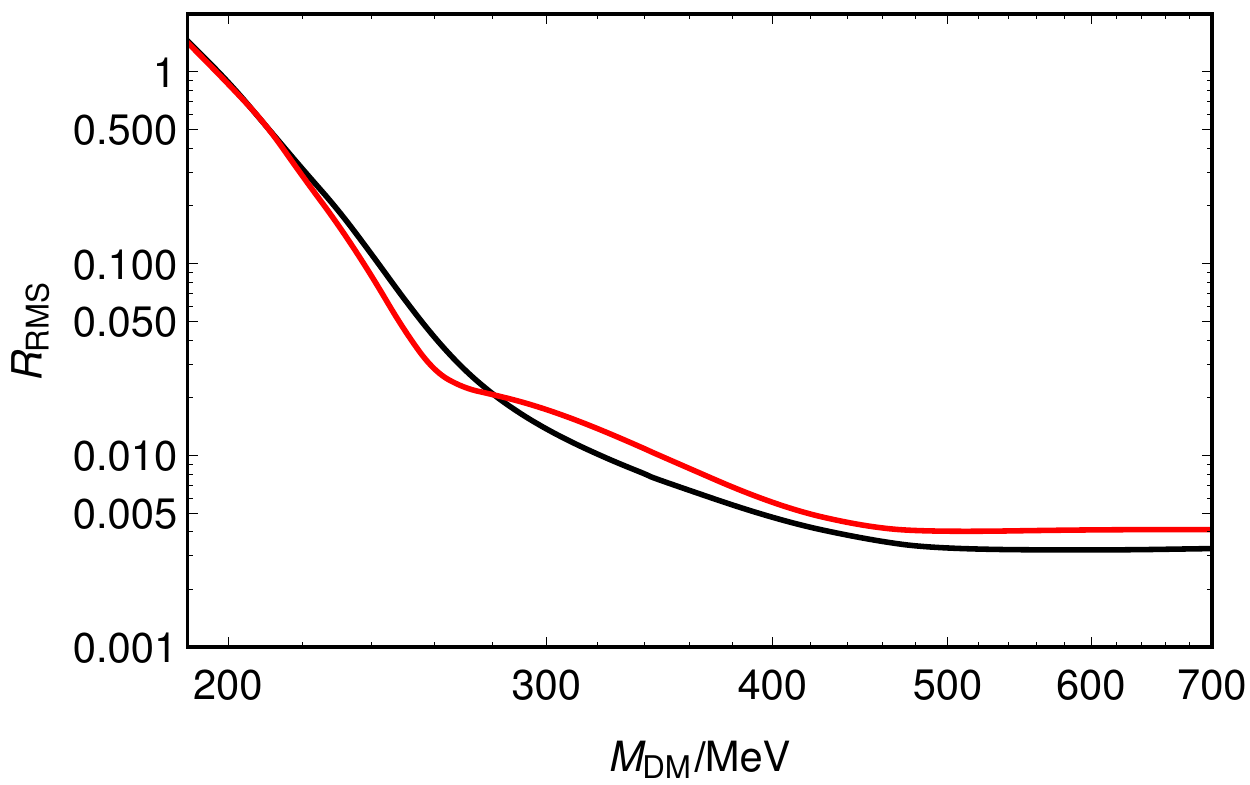}
\caption{Normalized RMS daily modulation in Germanium (left) and Silicon (right), for the intergrated Radon transform (black) and transverse Radon transform (red) as a function of the WIMP mass for the date of September 6, 2015.}
\label{fig:daily RMS}
\end{center}
\end{figure}

We define the normalized RMS daily modulation by
\be
R_{\rm RMS} = \sqrt{\frac{1}{\langle R \rangle^2\Delta t}\int_{\Delta t}(R(t)-\langle R\rangle)^2dt},
\ee
where $\langle R \rangle$ is the average event rate over the time interval $\Delta t$, and $R(t)$ is the event rate as a function of time. Figure~\ref{fig:daily RMS} shows $R_{\rm RMS}$ as a function of the WIMP mass, for Germanium and Silicon.

\subsection{Energy dependent interactions}

To demonstrate the behavior of the $q^2$-dependent scattering operators, we will focus here on the leading term in the $q^2$-expansion, the $q^2$, and the long-range force effective operator, $q^{-4}$.

\begin{figure}
\begin{center}
\includegraphics[width=0.32\textwidth]{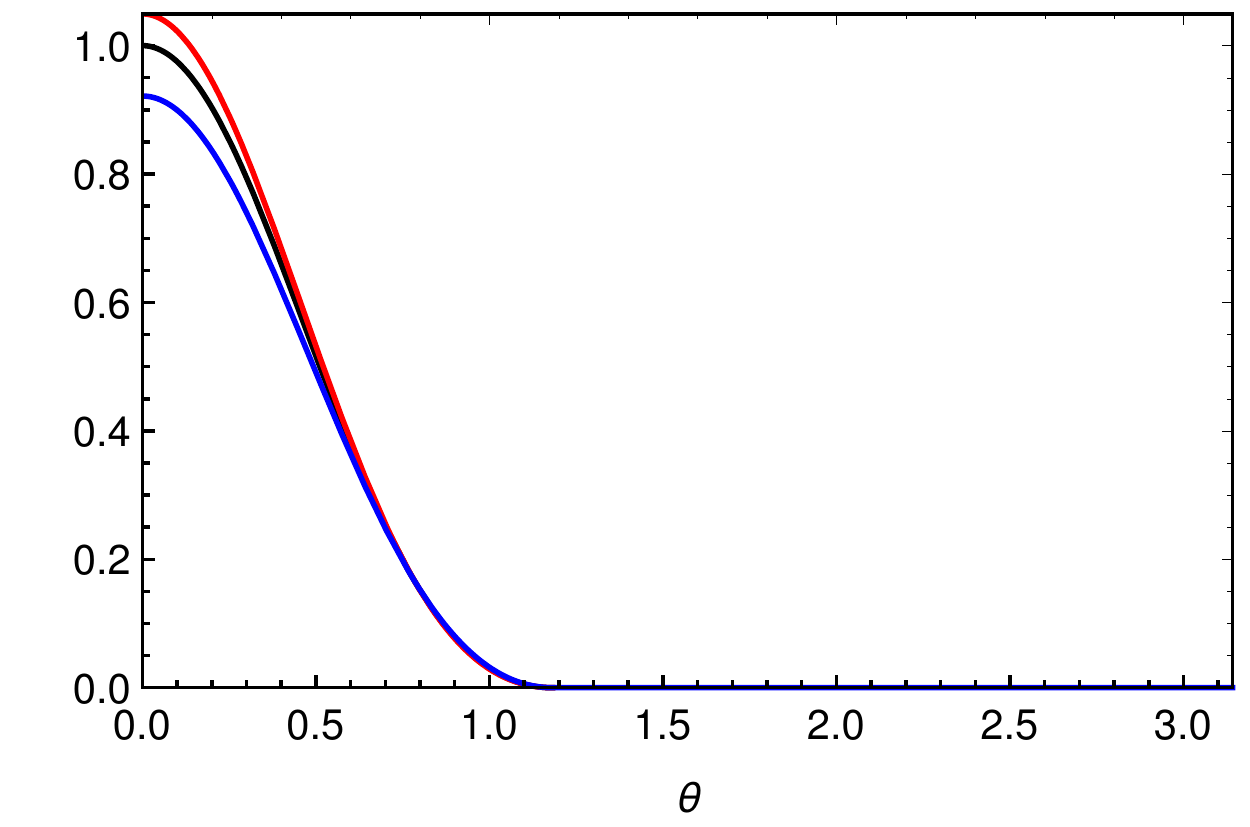}
\includegraphics[width=0.32\textwidth]{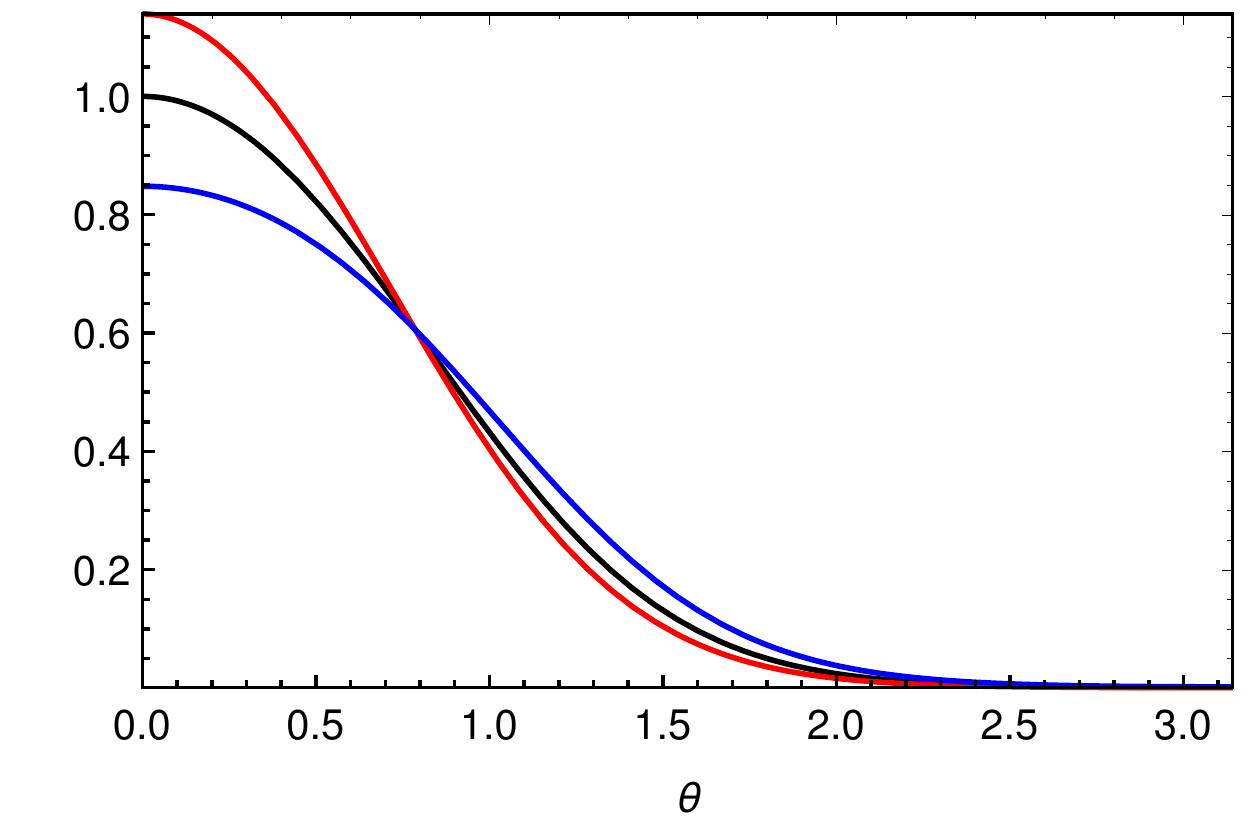}
\includegraphics[width=0.32\textwidth]{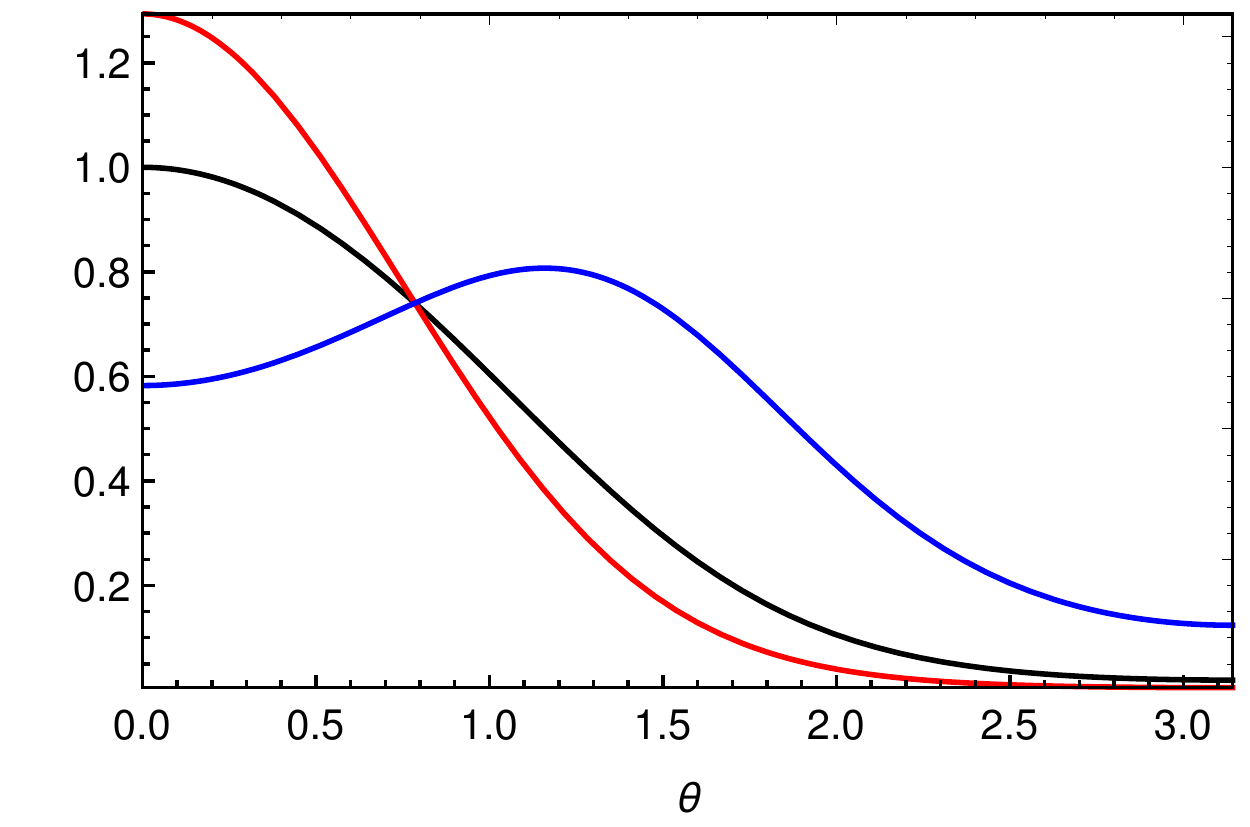}
\caption{Radon transform of the SHM, integrated from $E_{\rm min}=20$ eV to infinity, as a function of the angle $\theta$ between the average WIMP velocity and the recoil momentum, for $m_{DM}=0.4$ GeV (left), $m_{DM}=1$ GeV (center), $m_{DM}=5$ GeV (right). The black line corresponds to the operator $1$, the red line to the operator $q^2$ and the blue line to the operator $q^{-4}$.}
\label{RadontransformsLR}
\end{center}
\end{figure}

Figure \ref{RadontransformsLR} shows the integrated Radon transform (with $E_{\rm min}=20$ eV) as a function of the recoil direction for the operators $1$ (black), $q^2$ (red) and $q^{-4}$ (blue), for various values of the DM mass. Again we notice that the functions become similar to each other for small values of the DM mass.

\begin{figure}
\begin{center}
\includegraphics[width=0.45\textwidth]{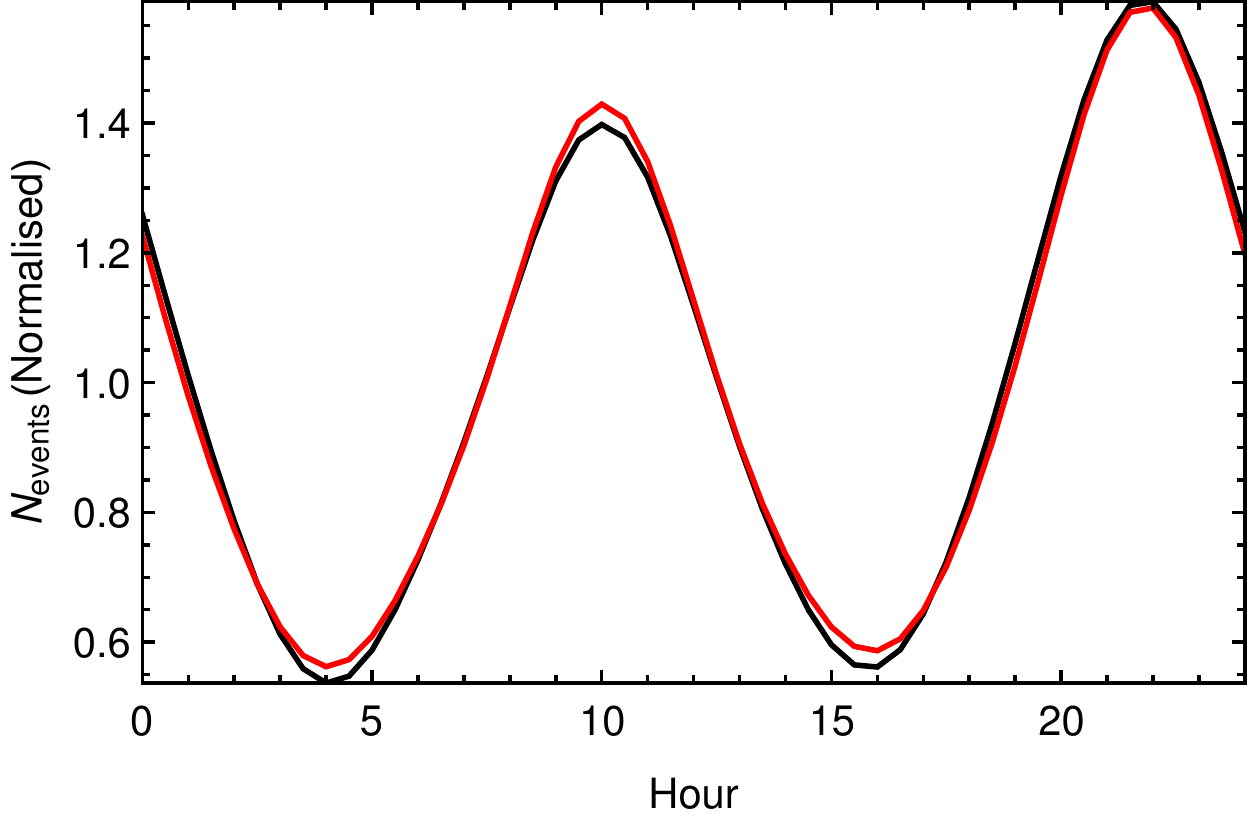}
\includegraphics[width=0.45\textwidth]{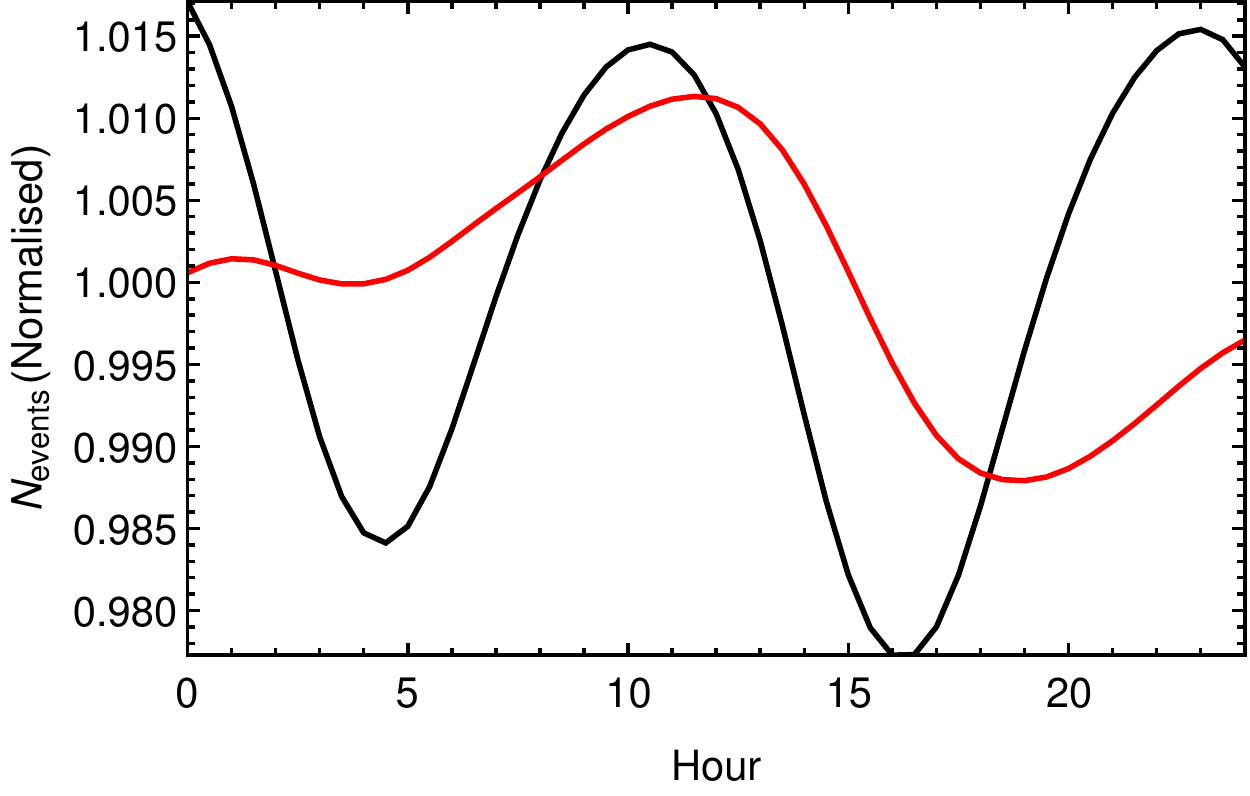}
\caption{Normalized event rate as a function of time on September 6, 2015, for $m_{DM}=0.3$ GeV (left) and $m_{DM}=0.4$ GeV (right). The black curve corresponds to the $q^{-4}$ operator and the red curve to the $q^2$ operator.}
\label{daily LR}
\end{center}
\end{figure}

\begin{figure}
\begin{center}
\includegraphics[width=0.75\textwidth]{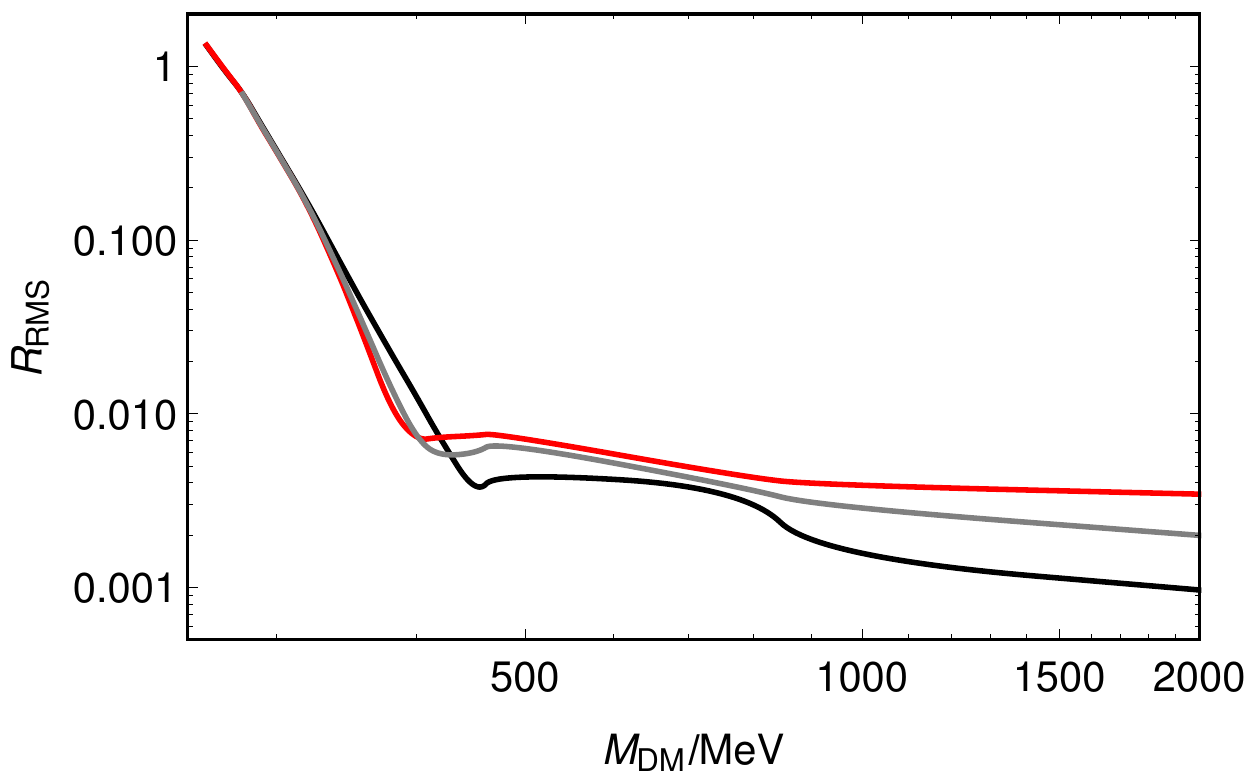}
\caption{Daily RMS modulation as a function of the DM mass for the operators $1$ (gray line), $q^2$ (red line) and $q^{-4}$ (black line).}
\label{RMS LR}
\end{center}
\end{figure}

This is also apparent in the daily rates, shown in figure \ref{daily LR}, for $m_{DM}=0.3$ GeV and $m_{DM}=0.4$ GeV wherein, the normalized event rates are nearly equal for the 0.3 GeV particle, but begin to deviate for larger values of the DM mass. The RMS modulation as a function of the DM mass is consistent with this observation, as shown in figure \ref{RMS LR}.

\begin{figure}
\begin{center}
\includegraphics[width = 0.32\textwidth]{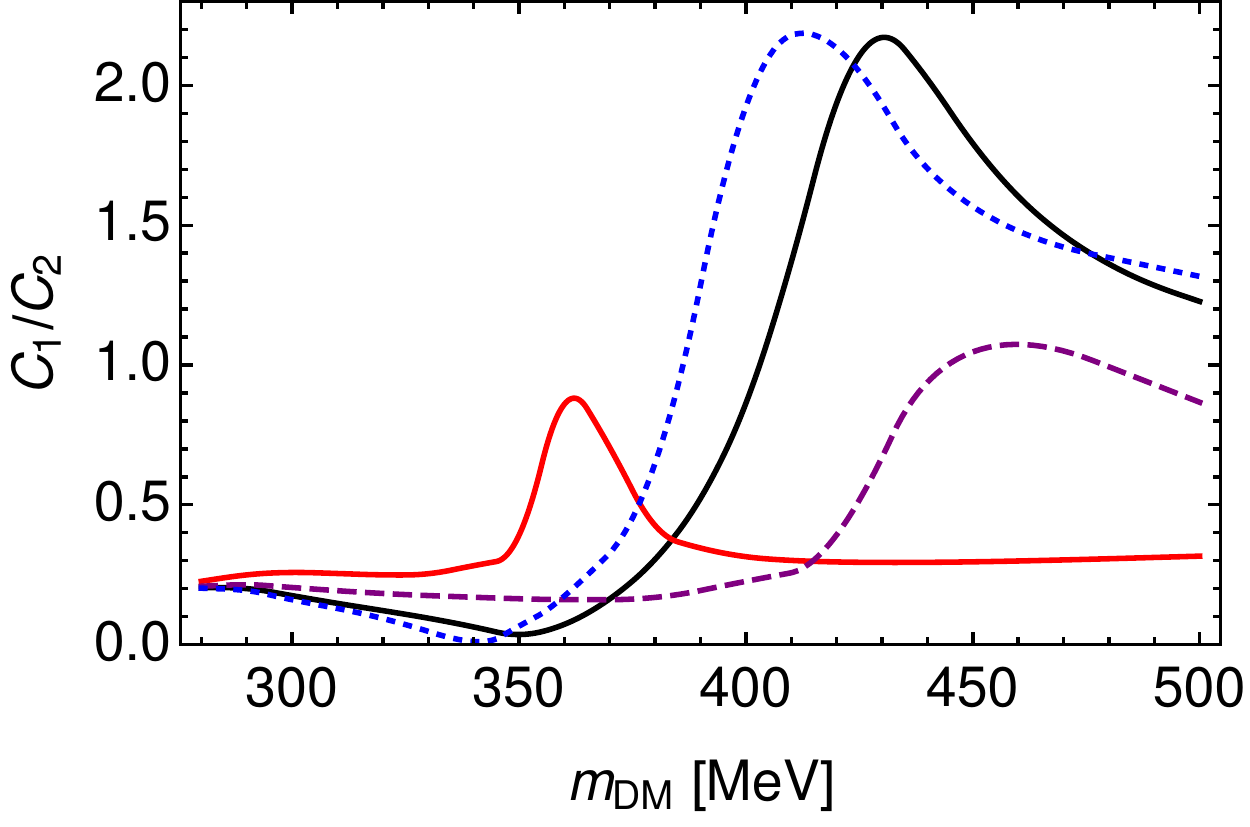}
\includegraphics[width = 0.32\textwidth]{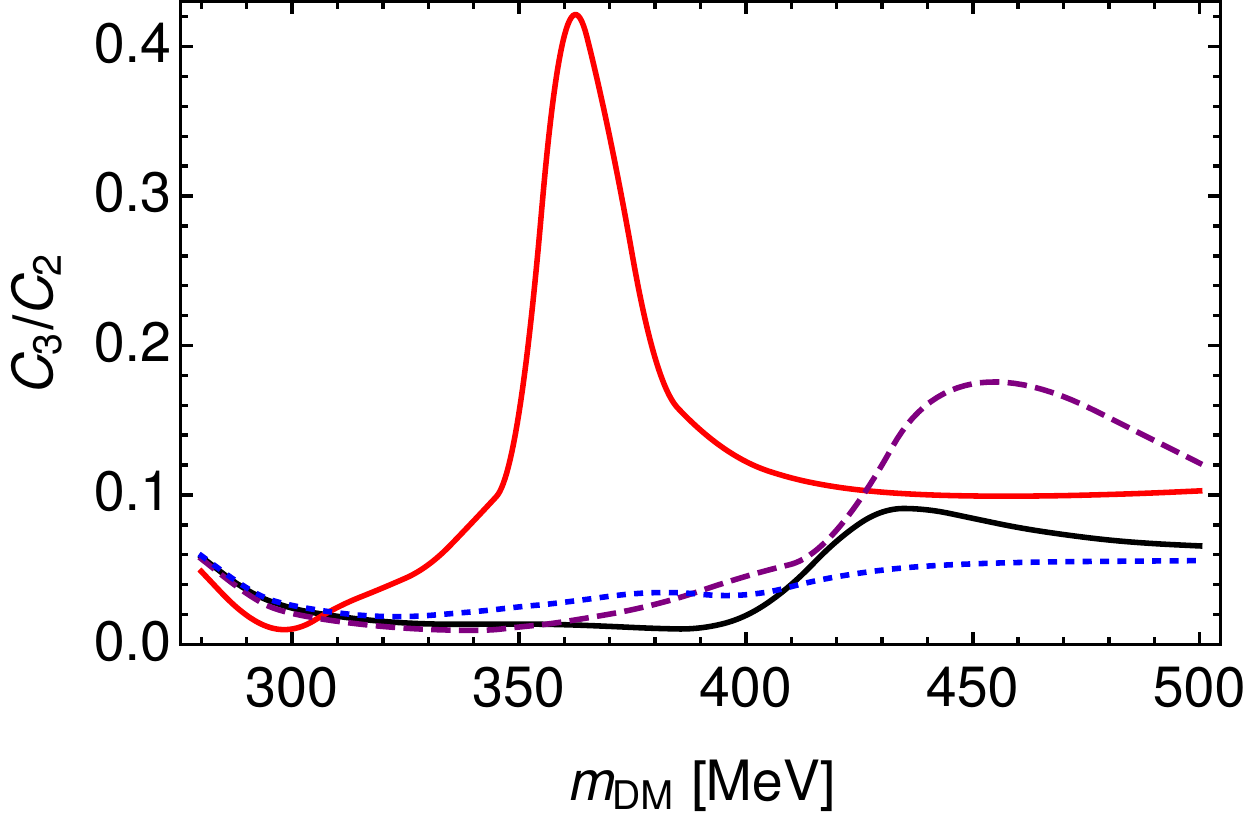}
\includegraphics[width = 0.32\textwidth]{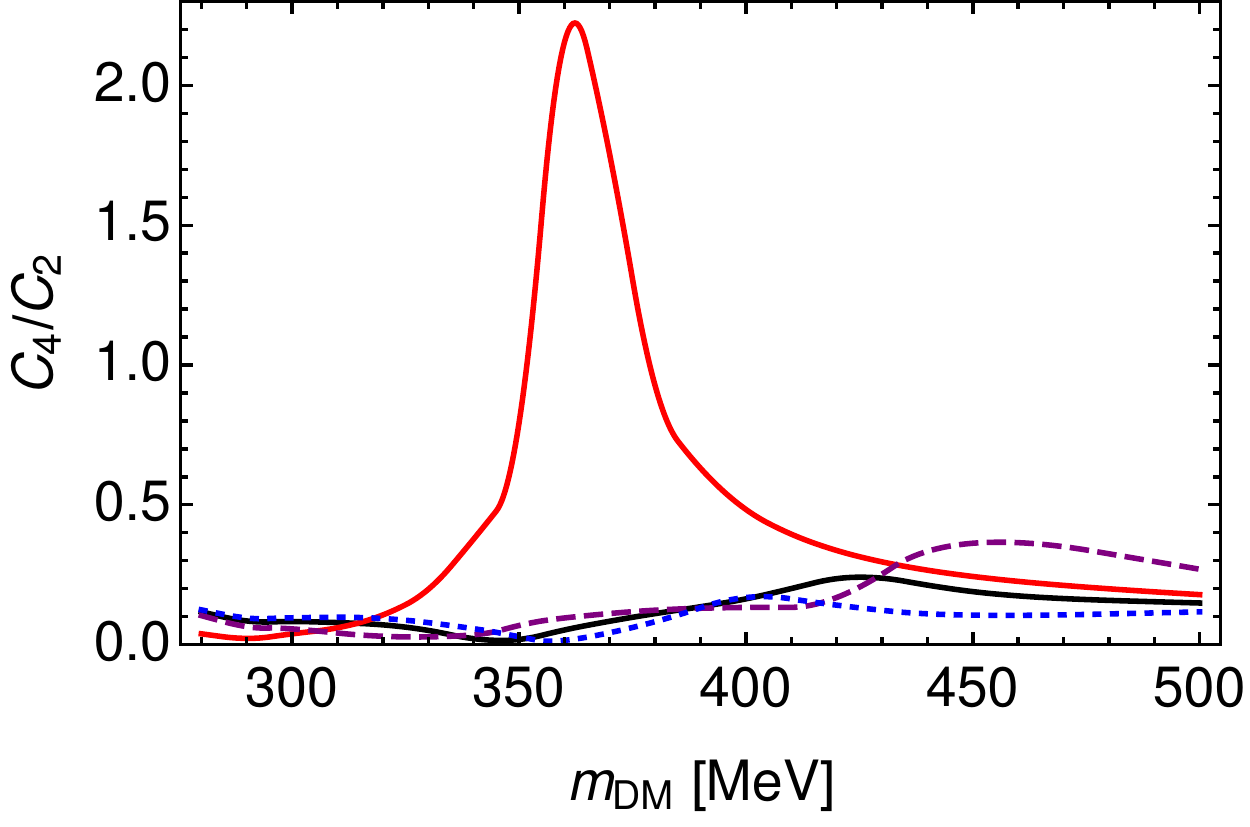}
\caption{The ratios of the Fourier-components $C_1/C_2$ (left), $C_3/C_2$ (center) and $C_4/C_2$ (right), for the $v^0$-interaction  (black line), $v_\perp^2$-interaction (red line), $q^{-4}$-interaction (purple dashed line) and $q^2$-interaction (blue dotted line), as a function of the DM mass.}
\label{fig:fouriermodes_all}
\end{center}
\end{figure}

Finally, in figure \ref{fig:fouriermodes_all} we show the same ratios of the Fourier-components as was shown in figure \ref{fig:fouriermodes} but here we also include the long-range interaction $q^{-4}$ and the $q^2$-interaction,  presented by the purple dashed and the blue dotted lines accordingly. We conclude that within the range $340\  {\rm MeV}\lesssim m_{\rm DM} \lesssim 450\ {\rm MeV}$ identified above, the long-range interaction can also be identified based on the ratios of the Fourier components. The case of the $q^2$-interaction shown in blue is more subtle, as the ratios of the Fourier-components for this interaction resemble those for the $v^0$-operator. However, within the most promising mass-window, these two can be additionally separated by the use of $C_1/C_2$-ratio.

\section{Summary and outlook}
\label{sec:conclusions}

We have considered dark matter scattering in the single-electron resolution ionization detectors wherein the quantum of electronic excitation 
$E_{\rm min}$ depends on the recoil direction. As established in~\cite{Kadribasic:2017obi}, the signal of dark matter scattering
in this case is detectable via the observation of diurnal modulation in expected event rates. We have extended the
analysis of~\cite{Kadribasic:2017obi} to cover the possible velocity and
energy dependencies of dark matter scattering on ordinary matter as implied
by the general low energy effective theory of dark matter.

We carried out the analysis using the Standard Halo Model for the DM velocity
distribution and our main finding is that for a given detector material there
is a range of sub-GeV masses, wherein the modulation signal is strong.
Furthermore, near the lower boundary of this mass range the shape of the modulation signal is practically independent on the nature of the underlying
DM interaction. However, above this lower boundary the signal develops a
strong dependence on the DM velocity and scattering energy, allowing for discrimination of different classes of interaction operators. At higher DM masses, towards one GeV, the overall amplitude of the modulation signal decreases and becomes less discernable.

In this study we have focused on germanium as the detector material, for which we identified the mass interval $340\  {\rm MeV}\lesssim m_{\rm DM} \lesssim 450\ {\rm MeV}$ as the most promising window where the type of the DM-SM scattering operator can be identified from the shape of the daily modulation signal. We have also performed a preliminary study on silicon, where we find qualitatively similar behavior, and identify the separation window as $250\  {\rm MeV}\lesssim m_{\rm DM} \lesssim 350\ {\rm MeV}$. With a selection of detector materials, it could thus be feasible to cover a larger range of DM masses, with multiple experiments having partly overlapping regions of sensitivity. The exploration of the directional dependence of the ionization energy threshold in a variety of materials is therefore strongly motivated.

There is a rising interest in the dark matter search community to develop very low threshold detectors ~\cite{contact_free} and single electron threshold has already been demonstrated in phonon mediated detectors ~\cite{CDMS_singleE}. A careful calibration of common semiconductors for dark matter detection down to the single electron-hole excitation level is necessary in order to interpret their results. Using mono-energetic neutron beams, such efforts are currently ongoing in various facilities and the results presented in this work can be verified in those experiments.

\appendix
\section{Radon transforms and their energy integrals}
\label{app:radontrans}
Consider the Radon transform of a function $f$, defined as
\be
\hat{f}(w,\hat{\bvec{w}})=\int\delta(w-\hat{\bvec{w}}\cdot\bvec{v})f(\bvec{v})d^3v,
\ee
and the transverse Radon transform
\be
\hat{f}^T(w,\hat{\bvec{w}})=\int\delta(w-\hat{\bvec{w}}\cdot\bvec{v})(\bvec{v}_\perp)^2 f(\bvec{v})d^3v,
\ee
where $\bvec{v}_\perp=\bvec{v}-(\bvec{v}\cdot\hat{\bvec{w}})\hat{\bvec{w}}$.

Choosing the $z$-axis parallel to the unit vector $\hat{\bvec{w}}$ and assuming isotropic velocity distribution $f(\bvec{v})=f(v)$, these can be expressed as integral over the amplitude $v$ only:
\beq
\hat{f}(w,\hat{\bvec{w}}) &=& 2\pi\int_w^\infty vf(v)dv, \\
\hat{f}^T(w,\hat{\bvec{w}})&=& 2\pi\int_w^\infty v(v^2-w^2)f(v)dv.
\eeq
The motion with respect to the galactic rest frame is taken into account via
the coordinate transformation $\bvec{v}\rightarrow \bvec{v}-\bvec{V}$, where $\bvec{V}$ is the velocity of the lab-frame with respect to the galactic rest frame. We follow the parametrization given in \cite{Bozorgnia:2011tk,Mayet:2016zxu} for $\bvec{V}(t)$.
In this case we have
\beq
\hat{f}(w,\hat{\bvec{w}}) &=& 2\pi\int_{w+V_z}^\infty vf(v)dv, \\
\hat{f}^T(w,\hat{\bvec{w}})&=& 2\pi\int_{w+V_z}^\infty v(v^2-2(w+V_z)V_z+V^2-w^2)f(v)dv,
\eeq
where $V_z=\bvec{V}\cdot\hat{\bvec{w}}$.
For a Maxwell distribution $f_M(v) = (2\pi\sigma_v^2)^{-\frac32}\exp(-v^2/(2\sigma_v^2)$ these are explicitly given as:
\beq
\hat{f}_M(w,\hat{\bvec{w}}) &=& \frac{1}{\sqrt{2\pi\sigma_v^2}}e^{-\frac{(w+V_z)^2}{2\sigma_v^2}}, \\
\hat{f}_M^T(w,\hat{\bvec{w}})&=& \frac{1}{\sqrt{2\pi\sigma_v^2}}(2\sigma_v^2+V^2-V_z^2)e^{-\frac{(w+V_z)^2}{2\sigma_v^2}}.
\eeq
The SHM distribution (\ref{eq:fSHM}) is $f_{\rm{SHM}}(v)=N_{\rm{e}}^{-1}f_M(v)\Theta(v_e-v)$, where $v_e$ is the escape velocity. Then the Radon transforms are given as:
\beq
\hat{f}_{\rm SHM}(w,\hat{\bvec{w}}) &=& \frac{N_{\rm{e}}^{-1}}{\sqrt{2\pi\sigma_v^2}}\left( e^{-\frac{(w+V_z)^2}{2\sigma_v^2}}-e^{-\frac{v_e^2}{2\sigma_v^2}}\right)\Theta(v_e-(w+V_z)), \label{eq:radonSHM}  \\
\hat{f}_{\rm SHM}^T(w,\hat{\bvec{w}}) &=& \frac{N_{\rm{e}}^{-1}}{\sqrt{2\pi\sigma_v^2}}\bigg( (2\sigma_v^2+V^2-V_z^2) e^{-\frac{(w+V_z)^2}{2\sigma_v^2}} \nonumber \\
&& -(2\sigma_v^2+V^2+v_e^2-2V_z^2-2wV_z-w^2)e^{-\frac{v_e^2}{2\sigma_v^2}}\bigg)\Theta(v_e-(w+V_z)).
\label{eq:TradonSHM}
\eeq
The directional event rate (\ref{eq:directional rate}) is obtained by integration over energy:
\beq
\int\limits_{E_{\rm min}}^{\infty}&\hat{f}_{\rm SHM}&(v_{\rm{min}},\hat{\bvec{q}})dE =N_{\rm{e}}^{-1}\Bigg( m V_z \text{erf}\left(\frac{\sqrt{E_{\rm min }}+\sqrt{m} V_z}{\sqrt{2m} \sigma_v}\right) + \sqrt{\frac{2}{\pi }} m \sigma_v  \left(e^{-\frac{2 V_z \sqrt{m E_{\rm min }}+E_{\rm min }+m V_z^2}{2
   m \sigma_v ^2}}-e^{-\frac{v_e^2}{2 \sigma_v ^2}}\right) \nonumber \\
&+& \frac{e^{-\frac{v_e^2}{2 \sigma_v ^2}}
   \left(E_{\rm min }-m (v_e-V_z)^2\right)}{\sqrt{2 \pi } \sigma_v }-m V_z
   \text{erf}\left(\frac{v_e}{\sqrt{2} \sigma_v }\right)\Bigg)\Theta\left(v_e-\sqrt{\frac{E_{\rm min}}{m}}-V_z\right),
\label{intRadon}
\eeq
where we now denote $V_z=\bvec{V}\cdot\hat{\bvec{q}}$ and $m=2\mu_{\rm DM,N}^2/m_{\rm N}$. For the transverse Radon transform the integral over energy reads:
\beq
\int\limits_{E_{\rm min}}^{\infty}&&\hat{f}_{\rm SHM}^T(v_{\rm{min}},\hat{\bvec{q}})dE = \frac{N_{\rm{e}}^{-1}}{\sqrt{2 \pi } \sigma_v }\Bigg( m \sigma_v  \left(2 \sigma_v ^2+V^2-V_z^2\right) \Bigg[ \sqrt{2 \pi } V_z \left(\text{erf}\left(\frac{\sqrt{E_{\rm min }}+\sqrt{m} V_z}{\sqrt{2m} \sigma_v }\right)-\text{erf}\left(\frac{v_e}{\sqrt{2} \sigma_v }\right)\right) \nonumber \\
&& +2 \sigma_v  \left(e^{-\frac{2 V_z \sqrt{m E_{\rm min }}+E_{\rm min }+m V_z^2}{2 m \sigma_v ^2}}-e^{-\frac{v_e^2}{2 \sigma_v^2}}\right)\Bigg] +\frac{1}{6 m}e^{-\frac{v_e^2}{2 \sigma_v ^2}} \Bigg[ 6 m E_{\rm min } \left(2 \sigma_v^2+V^2+v_e^2-2 V_z^2\right) \nonumber \\
&& -8 V_z \sqrt{m E_{\rm min }^3}-3 E_{\rm min }^2-m^2(v_e-V_z)^2 \left(12 \sigma_v^2+6V^2+3v_e^2-2v_e V_z-7V_z^2\right) \Bigg] \Bigg) \nonumber \\
&& \times \Theta\left(v_e-\sqrt{\frac{E_{\rm min}}{m}}-V_z\right).
\label{intTRadon}
\eeq
For the $q^2$ operator, the Radon transform (\ref{eq:radonSHM}) must be multiplied by $q^2$ before taking the integral over energy. The result is
\beq
\int\limits_{E_{\rm min}}^{\infty}&q^2\hat{f}_{\rm SHM}&(v_{\rm{min}},\hat{\bvec{q}})dE =N_{\rm{e}}^{-1}\Bigg(
\sqrt{\frac{2}{\pi }} m m_{\rm N} \Bigg[ \sqrt{2 \pi } m V_z \left(3 \sigma_v^2+V_z^2\right) {\rm erf}\left(\frac{\sqrt{E_{\rm min}}+\sqrt{m}
   V_z}{\sqrt{2} \sqrt{m} \sigma_v }\right) \nonumber \\
&&+2 \sigma_v  e^{-\frac{2 V_z \sqrt{m E_{\rm min }}+E_{\rm min }+m V_z^2}{2 m \sigma_v ^2}}
   \left(-V_z\sqrt{m E_{\rm min }}+E_{\rm min }+m \left(2 \sigma_v ^2+V_z^2\right)\right) \nonumber \\
&&-\sqrt{2 \pi } m V_z \left(3 \sigma_v
   ^2+V_z^2\right) \text{erf}\left(\frac{v_e}{\sqrt{2} \sigma_v }\right)-2 m \sigma_v  e^{-\frac{v_e^2}{2 \sigma_v ^2}} \left(2 \sigma_v
   ^2+v_e^2-3 v_e V_z+3 V_z^2\right) \Bigg] \nonumber\\
&&+\frac{m_{\rm N} e^{-\frac{v_e^2}{2 \sigma_v ^2}} \left(E_{\rm min }^2-m^2
   (v_e-V_z)^4\right)}{\sqrt{2 \pi } \sigma_v } \Bigg) \Theta\left(v_e-\sqrt{\frac{E_{\rm min}}{m}}-V_z\right).
\eeq
For the integral $\int\limits_{E_{\rm min}}^{\infty}q^{-4}\hat{f}_{\rm SHM}(v_{\rm{min}},\hat{\bvec{q}})dE$ we find no analytic expression, and therefore perform the integral over energy numerically in this case.

\end{document}